\documentclass[twocolumn]{svjour3}         
\smartqed  
\usepackage{graphicx}
\newcommand{\cd}{\makebox[0.08cm]{$\cdot$}}

\newcommand{\sla}{\not\!}
\begin{document}

\title{Nonperturbative renormalization in light-front dynamics and applications
\thanks{Relativistic Description of Two- and Three-Body Systems in Nuclear Physics", ECT*, October 19-23 2009}}

\author{J.-F. Mathiot           \and
        A.V. Smirnov            \and
        N.A. Tsirova              \and
        V.A. Karmanov           }

\institute{
          J.-F. Mathiot  \at
              Clermont Universit\'e, Laboratoire de Physique Corpusculaire, \\ BP10448, F-63000 Clermont-Ferrand, France\\
             \email{mathiot@clermont.in2p3.fr}
           \and
          A.V. Smirnov  \at
              Lebedev Physical Institute, Leninsky Prospekt 53, \\ 119991 Moscow, Russia\\
              \email{avs44@rambler.ru}
           \and
          N.A. Tsirova  \at
              Clermont Universit\'e, Laboratoire de Physique Corpusculaire, \\ BP10448, F-63000 Clermont-Ferrand, France\\
             \email{tsirova@clermont.in2p3.fr}
             \and
            V.A. Karmanov \at
             Lebedev Physical Institute, Leninsky Prospekt 53, \\ 119991 Moscow, Russia\\
             \email{karmanov@sci.lebedev.ru}
}
\maketitle
\begin{abstract}
We present a general framework to calculate the properties of
relativistic compound systems from the knowledge of an elementary
Hamiltonian. Our framework provides a well-controlled
nonperturbative calculational scheme which can be systematically
improved. The state vector of a physical system is
calculated in light-front dynamics. From the general properties of
this form of dynamics, the state vector can be further decomposed
in well-defined Fock components. In order to control the
convergence of this expansion, we advocate the use of the
covariant formulation of light-front dynamics. In this
formulation, the state vector is projected on an arbitrary
light-front plane $\omega \cd x=0$ defined by a light-like
four-vector $\omega$. This enables us to control any violation of
rotational invariance due to the truncation of the Fock expansion.
We then present a general nonperturbative renormalization scheme
in order to avoid field-theoretical divergences which may remain
uncancelled due to this truncation. This general framework has been
applied to a large variety of models. As a starting point, we
consider QED for the two-body Fock space truncation and calculate
the anomalous magnetic moment of the electron. We show that it
coincides, in this approximation, with the well-known
Schwinger term. Then we investigate the properties of a purely
scalar system in the three-body approximation, where we highlight
the role of antiparticle degrees of freedom. As a non-trivial
example of our framework, we calculate the structure of  a
physical fermion in the Yukawa model, for the three-body Fock
space truncation (but still without antifermion contributions). We
finally show why our approach is also well-suited to describe
effective field theories like chiral perturbation theory in the
baryonic sector.

\end{abstract}

\section{Light-front dynamics in few-body systems and field theory} \label{intro}
The understanding of hadron properties from an underlying
Lagrangian or Hamiltonian is a major issue in nuclear and particle
physics. It demands both a relativistic framework to deal with
quasi-massless particles (the pion, up and down quarks, etc.) or
with high momentum and high energy experiments, and a
nonperturbative framework. The latter is mandatory in order to
calculate for instance the mass of a bound state from the pole of
the scattering amplitude or from an eigenstate equation.

One may already gain some physical insights from nonrelativistic
studies like the nonrelativistic constituent quark model in
particle physics, or the study of few-nucleon systems, based on
the nonrelativistic nucleon-nucleon or three-nucleon potentials,
in nuclear physics. This has given rise to numerous studies in the
last 40 years.

The extension to relativistic calculations may either rely on the
use of relativistic equations like the four-dimensional
Bethe-Salpeter equation (including its various three-dimensional
quasipotential reductions) and the Dyson-Schwinger equation, or on
Hamiltonian dynamics, using one of its three forms proposed by
Dirac in 1949~\cite{Dirac}. We shall follow in this review the
path pointed out by Dirac and choose light-front dynamics (LFD) as
a basis of our approach.

\subsection{Few-body relativistic systems}
A natural testing ground for the use of LFD is the study of
few-body systems. The properties of this particular form of
dynamics are indeed very well-suited to make a tight connection
with nonrelativistic considerations.

In the standard form of LFD, the state vector of a physical system
is defined not at a fixed moment of time but on the light-front
plane given by the equation $t+\frac{z}{c}=const$. The
nonrelativistic limit reached by taking $c \to \infty$ leads thus
naturally to the ordinary equal-time formulation $t=const$, giving
rise, in particular, to the Schr\"odinger equation for the
nonrelativistic wave function.

One can always decompose the state vector in
Fock components. Since the physical vacuum is trivial in LFD, i.e.
it coincides with the vacuum for not interacting particles, this
decomposition does not include the vacuum fluctuations but
contains the physical states only. Few-body systems are
represented just by the first few components of this expansion.
Moreover, each Fock component has a probabilistic interpretation
similar to that of the nonrelativistic wave function.

In this formulation, the elementary kernel of the eigenstate
equation defining the state vector of the physical system is
calculated in light-front time-ordered perturbation theory. All
four-momenta are on the mass shell, while all intermediate states
are off the energy shell. The eigenstate equation to be solved is
then three-dimensional, in direct analogy with the nonrelativistic
Schr\"odinger equation.

According to the Dirac's classification, the ten generators of the
Poincar\'e group, given by space-time translations (four
generators), space rotations (three generators), and Lorentz
boosts (three generators), can be separated into kinematical and
dynamical operators. The kinematical operators leave the
light-front plane invariant and are independent of dynamics, i.e.
of the interaction Hamiltonian of the system, while the dynamical
ones change the light-front position and depend therefore on the
interaction. Among the kinematical operators, one finds, in LFD,
the boost along the $z$ axis. This property is of particular
interest when one calculates electromagnetic observables at high
momentum transfer, since once one knows the state vector in one
reference frame, it is easy to calculate it in any other frame.

One has of course to pay some price for that. The spatial
rotations in the $xz$ and $yz$ planes become dynamical, in
contrast to the case of equal-time dynamics. This is a direct
consequence of the violation of rotational invariance caused by
the non-invariant definition of the light-front plane orientation.
This violation should be kept under control.

\paragraph{Control of rotational invariance.}
While rotational invariance should be recovered automatically in
any exact calculation, this is not {\em a priori} the case if the
Fock expansion is truncated. The control of the violation of
rotational symmetry is very difficult in practice, when using the
standard form of LFD. To avoid such an unpleasant feature of the
latter, we shall use below the covariant formulation of LFD
(CLFD)~\cite{karm76,cdkm}, which provides a simple, practical, and
very powerful tool in order to describe physical systems as well
as their electromagnetic amplitudes. In this formulation, the
state vector is defined on the plane characterized by  the
invariant equation $\omega \cd x=0$, where $\omega$ is an
arbitrary light-like ($\omega^2=0$) four-vector. The  standard LFD
on the plane $t+\frac{z}{c}=0$ is recovered by considering the
particular choice $\omega=(1,0,0,-1)$. The covariance of our
approach relies on the invariance of the light-front plane
equation under any Lorentz transformation of both $\omega$ and
$x$. This implies in particular that $\omega$ cannot be kept the
same in any reference frame,  as it is the case in the standard
formulation of LFD.

There is of course equivalence, in principle, between the
standard and covariant forms of LFD in any exact calculation.
Calculated physical observables must coincide in both approaches,
though their derivation in CLFD in most cases is much simpler and
more transparent. Indeed, the relation between CLFD and standard
LFD reminds that between the Feynman graph technique and
old-fashioned perturbation theory.

In approximate calculations however, CLFD has a definite advantage
in the sense that it enables a direct handle on the contributions
which violate rotational invariance. These ones depend explicitly
on the orientation of the light-front surface (i.e. on $\omega$)
and can thus be separated covariantly from true physical
contributions. This is of particular interest when one considers
for instance electromagnetic observables in few-body
systems~\cite{cdkm}, or field theory on the light front.

\subsection{Light-front field theory} \label{deltam}
The interest to the application of LFD to field-theoretical
problems is also very old. It originates from the study of deep
inelastic scattering experiments. Indeed, it is easy to see that
in this kinematical domain, all events are close to the
light-front plane. This legitimates the use of the infinite
momentum frame in the calculation of high energy
observables~\cite{IMF}. The calculation is done in perturbation
theory. The extension to QED is natural, since perturbation theory
is also applicable in that case~\cite{brs}.

The use of LFD in field theory is important in order to extend
these calculations to the nonperturbative domain~\cite{bpp}. As
already mentioned, the structure of the vacuum in LFD enables a
well-defined expansion of the state vector in Fock components. In
field theory, this expansion is in principle infinite. From a
practical point of view, it should be truncated to a limited
number of components. We thus have to worry about the convergence
of this expansion, and the various ways to speed it up if one wants to be able to
make meaningful predictions.

The truncation of the Fock expansion induces however two
pernicious features in the study of field theory on the light
front. The first one, which we have already addressed in the
discussion of few-body systems, is the violation of rotational
invariance due to the particular choice of the orientation of the
light-front plane. The second one is the appearance of uncancelled
divergences which calls for an appropriate renormalization scheme.

\paragraph{Appropriate renormalization scheme.}
The truncation of the Fock expansion complicates the
renormalization procedure, in contrast to that in standard
perturbation theory. Indeed,  the full cancellation of
field-theoretical divergences which appear in a given Fock sector
requires taking into account contributions from other sectors. If
even a part of the latter is beyond our approximation, some
divergences may leave uncancelled.

\begin{figure}[h]
\begin{center}
\includegraphics[width=20pc]{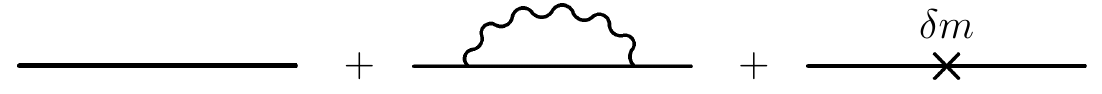}
\caption{Renormalization of the fermion propagator in the second
order of perturbation theory. The fermion mass counterterm is
denoted by $\delta m$. \label{self}}
\end{center}
\end{figure}

For instance, looking at Fig.~\ref{self} for the calculation of
the fermion propagator in the second order of perturbation theory,
one immediately realizes that the cancellation of divergences
between the self-energy contribution (of 2nd
order in the Fock decomposition) and the fermion mass counterterm
(of 1st order one) involves two different Fock sectors.

This means that, as a necessary condition for the cancellation of
divergences, any mass counterterm  should be associated  with the
number of particles present (or ``in flight'') in a given Fock
sector. In other words, all mass counterterms must depend on the
Fock sector under consideration, as advocated first in Ref.~\cite{wp}.
This is also true for the renormalization of the bare coupling
constant.

The presence of uncancelled divergences reflects itself in
possible dependence of approximately calculated observables on the
regularization parameters (e.~g., cutoffs). In other words,
calculated physical observables are not anymore scale invariant. This
prevents to make any physical predictions if we cannot control the
renormalization procedure in one way or another. We have developed
an appropriate renormalization procedure --- the so-called Fock
sector dependent renormalization (FSDR) scheme --- in order to
keep the cancellation of field-theoretical divergences under
permanent control. This scheme relies also directly on CLFD in
order to define the renormalization conditions imposed on true
physical observables~\cite{kms_08}.

We shall detail in the following the main features of CLFD, as
well as FSDR, and apply our general framework to various physical
systems.

\section{Covariant formulation of light-front dynamics} \label{CLFD}

The state vector, $\phi_\omega^{J\sigma}(p)$, of a compound system
corresponds to definite values for its mass $M$, its four-momentum
$p$, and its total angular momentum $J$ with the projection
$\sigma$ onto the $z$ axis in the rest frame, i.e., the state
vector forms a representation of the Poincar\'e group. It
satisfies the following eigenstate equations:
\begin{eqnarray}\label{kt14}
\hat{P}_{\rho}\ \phi_\omega^{J\sigma}(p)&=&p_{\rho}\ \phi_\omega^{J\sigma}(p),\\
\label{kt15}
\hat{P}^{2}\ \phi_\omega^{J\sigma}(p)&=&M^2\ \phi_\omega^{J\sigma}(p),\\
\label{kt16}
\hat{S}^{2}\ \phi_\omega^{J\sigma}(p)&=&-M^2\ J(J+1)\ \phi_\omega^{J\sigma}(p),\\
\label{kt17} \hat{S}_{3}\ \phi_\omega^{J\sigma}(p)&=&M\
\sigma\phi_\omega^{J\sigma}(p),
\end{eqnarray}
where $\hat{P}_{\rho}$ is the four-momentum operator,
$\hat{S}_{\rho}$ is the Pauli-Lubanski vector
\begin{equation}\label{kt18}
\hat{S}_{\rho}= \frac{1}{2}\epsilon_{\rho\nu\alpha\beta} \
\hat{P}^{\nu}\ \hat{J}^{\alpha\beta},
\end{equation}
and $\hat{J}$ is the four-dimensional angular momentum
operator which is represented as a sum of the free and interaction parts:
\begin{equation}
\label{kt2} \hat{J}_{\rho\nu}=\hat{J}^{(0)}_{\rho\nu}
+\hat{J}^{int}_{\rho\nu}.
\end{equation}
In terms of the interaction Hamiltonian $H^{int}(x)$ we have
\begin{equation}\label{kt5}
 \hat{J}^{int}_{\rho\nu}=\int H^{int}(x)(x_{\rho}\omega_{\nu} -x_{\nu}
\omega_{\rho}) \delta(\omega\cd x)\ d^4x.
\end{equation}

Similarly to $\hat{J}$, the momentum operator also can be split
into the free and interaction parts:
\begin{equation} \label{pshr}
\hat{P}_{\rho}=\hat{P}^{(0)}_{\rho} +\hat{P}^{int}_{\rho},
\end{equation}
with
\begin{equation}\label{pintham}
\hat{P}^{int}_{\rho}=\omega_{\rho}\int
H^{int}(x)\,\delta(\omega\cd x)\,d^4x.
\end{equation}

From the general transformation properties of both the  state
vector and the light-front plane, it follows~\cite{k82} that
\begin{equation}\label{kt12}
\hat{J}^{int}_{\rho\nu} \ \phi_\omega^{J\sigma}(p)=
\hat{L}_{\rho\nu}(\omega)\phi_\omega^{J\sigma}(p),
\end{equation}
where
\begin{equation}\label{kt13}
\hat{L}_{\rho\nu}(\omega) =i\left(\omega_{\rho}
\frac{\partial}{\partial\omega^{\nu}} -\omega_{\nu}
\frac{\partial}{\partial\omega^{\rho}}\right).
\end{equation}
The equation~(\ref{kt12}) is called the {\it angular condition}.
We can use it in order to replace the operator
$\hat{J}^{int}_{\rho\nu}$ entering into Eq.~(\ref{kt18}) by
$\hat{L}_{\rho\nu}(\omega)$. Introducing the notations
\begin{eqnarray}\label{kt19}
\hat{M}_{\rho\nu} &=&\hat{J}^{(0)}_{\rho\nu} +\hat{L}_{\rho\nu}(\omega),\\
\hat{W}_{\rho}&=&
\frac{1}{2}\epsilon_{\rho\nu\alpha\beta} \ \hat{P}^{\nu}\
\hat{M}^{\alpha\beta},\label{kt20}
\end{eqnarray}
we obtain, instead of Eqs.~(\ref{kt16}) and~(\ref{kt17}),
\begin{eqnarray}\label{kt21}
\hat{W}^{2}\phi_\omega^{J\sigma}(p)&=&-M^2J(J+1)\ \phi_\omega^{J\sigma}(p),\\
\label{kt22} \hat{W}_{3}\ \phi_\omega^{J\sigma}(p)&=&M\ \sigma\
\phi_\omega^{J\sigma}(p).
\end{eqnarray}
{\it These equations do not contain the interaction Hamiltonian,
once} $\phi_\omega^{J\sigma}$ {\it satisfies Eqs.}~(\ref{kt14})
{\it and}~(\ref{kt15}). The construction of the state vector of a
physical system with definite total angular momentum becomes
therefore {\it a purely kinematical problem}. Indeed, the
transformation properties of the state vector under rotations of
the coordinate system are fully determined by its total angular
momentum, while the dynamical part of the latter is separated out
by means of the angular condition. The dynamical dependence of the
state vector on the light-front plane orientation turns now into
its explicit dependence on the four-vector $\omega$~\cite{cdkm}.
Such a separation, in a covariant way, of kinematical and
dynamical transformations is a definite advantage of CLFD, as
compared to standard LFD.

%
\subsection{General Fock decomposition of the state vector}
\begin{figure}[ht!]
\begin{center}
\includegraphics[width=14pc]{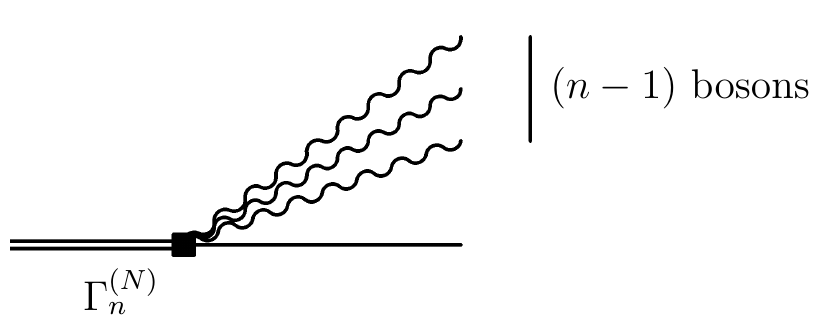}
\end{center}
 \caption{$n$-body vertex function
for a Fock space truncation of order $N$, for a physical fermion
(double straight line) made of a constituent fermion (single
straight line) coupled to bosons(wavy lines).}
 \label{gamman}
\end{figure}

According to the general properties of LFD, we decompose the state vector of a physical
system in Fock sectors. We have
\begin{eqnarray}
\phi(p)&=&
\sum_{n=1}^{ \infty} \int dD_n \phi_n(k_1,\ldots,k_n;p) \nonumber \\
&\times& \delta^4(k_1+\ldots +k_n-p-\omega \tau_n) \left\vert n
\right>,\label{Fock}
\end{eqnarray}
where $\left\vert n \right>$ is the state containing $n$ free
particles with the four-momenta $k_1,\ldots,k_n$ and $\phi_n$'s
are relativistic $n$-body wave functions or the so-called Fock
components. Here and below we will omit, for shortness, all spin
indices in the notation of the state vector. Note the particular
overall momentum conservation law given by the $\delta$-function.
It follows from the general transformation properties of the
light-front plane $\omega \cd x=0$ under four-dimensional
translations. The quantity $\tau_n$ is a measure of how far the
$n$-body system is off the energy shell\footnote{The term "off the
energy shell" is borrowed from the equal-time dynamics where the
spatial components of the four-momenta are always conserved, but
the energies of intermediate states are not equal to the incoming
energy.}. It is completely determined by this conservation law and
the on-mass-shell condition for each individual particle momentum.
We get
\begin{equation}
\label{tau}
2 \omega \cd p \ \tau_n=(s_n-M^2),
\end{equation}
where
\begin{equation} \label{sn}
s_n=(k_1+\ldots +k_n)^2.
\end{equation}
The phase space volume element is represented schematically by $d
D_n$. It involves integrations over the components of the
constituent four-momenta $d^4k_i$ and over $d\tau_n$ in infinite
limits. The state $\vert n\rangle$ can be written as
\begin{equation}
\label{freefield}
\left\vert n \right> \equiv d^\dagger (k_1) d^\dagger(k_2) \ldots
d^\dagger(k_{n}) \left\vert 0 \right>,
\end{equation}
where $d^\dagger$  is a generic notation for free particle
creation operators. To completely determine the state vector, we
normalize it according to
\begin{equation}
\label{normep} \phi(p') ^\dagger \phi(p) = 2 p_0 \delta^{(3)}({\bf
p'}-{\bf p}).
\end{equation}
It is convenient to introduce, instead of the wave functions
$\phi_n$, the vertex functions $\Gamma_n$ (which we will also
refer to as Fock components), defined by
\begin{equation}
\label{Gn}
\Gamma_n =(s_n-M^2)\phi_n \equiv 2
\omega \cd p \ \tau_n \phi_n.
\end{equation}
In the particular case of a fermion coupled to bosons, it is
convenient to extract from $\Gamma_n$ the fermion bispinors, and
make the replacement
\begin{equation}
\Gamma_n \to \bar u(k_1) \Gamma_n u(p),
\end{equation}
where $k_1$ is the four-momentum of the constituent fermion. When
Fock space is truncated, it is necessary to keep track of the
order of truncation $N$ (i.e. the maximal number of particles
admitted in the Fock sectors) in the calculation of
the vertex function. For this purpose we will use the notation
$\Gamma_n^{(N)}$ for the $n$-body vertex function. In the LFD
graph technique, it is represented by a $(n+1)$-leg vertex with one
incoming double line corresponding to the physical
state and $n$ outgoing single lines corresponding to
constituents. By its spin structure and transformation properties
it is completely analogous to a $n$-body wave function. As an
example, we show  such a vertex in Fig.~\ref{gamman} for the case
of a physical fermion state composed from one constituent fermion
and $(n-1)$ bosons. For simplicity, we shall use below the same
notation for the vertex functions of both fermion and boson
physical states.

With the decomposition~(\ref{Fock}), the normalization
condition~(\ref{normep}) writes
\begin{equation} \label{Inor}
 \sum_{n=1}^\infty I_n=1,
\end{equation}
where $I_n$ is the contribution of the $n$-body Fock sector to the
full norm of the state vector, equal to 1. The explicit formulas
for $I_n$ in terms of the vertex functions for some important
particular cases can be found in~\cite{kms_08}.

\subsection{Eigenstate equation}
The system of coupled equations for the Fock components of the
state vector can be obtained from Eq.~(\ref{kt15}) by substituting
there the Fock decomposition~(\ref{Fock}) and calculating the
matrix elements of the operator $\hat{P}^2$ in Fock space. With
the expressions~(\ref{pshr}) and~(\ref{pintham}), we get the
eigenstate equation~\cite{bckm}:
\begin{eqnarray}\label{eq1b}
2(\omega\cd p)\int \tilde{H}^{int}(\omega\tau)\frac{d\tau}{2\pi}
\phi(p)= -\left[\left(\hat{P}^{(0)}\right)^2-M^{2}\right]\phi(p),\nonumber \\
\end{eqnarray}
where $\tilde H^{int}$ is the interaction Hamiltonian in momentum
space:
\begin{equation}
\label{hamG} \tilde{H}^{int}(\omega\tau)=\int
H^{int}(x)e^{-i(\omega\cd x)\tau}d^4x.
\end{equation}

Using the general momentum conservation law in Eq.~(\ref{Fock}),
we conclude that the operator in the square brackets on the
right-hand side of Eq.~(\ref{eq1b}) simply multiplies each Fock
component of the state vector by the factor
$s_n-M^{2}\equiv2(\omega\cd p)\tau_n$. It is therefore convenient
to introduce the notation
\begin{equation}
{\cal G}(p)=2(\omega\cd p)\hat{\tau}\phi(p),
\end{equation}
where $\hat \tau $ is the operator which, acting on a given
component $\phi_{n}$ of $\phi(p)$, gives $\tau_n
\phi_{n}$. ${\cal G}(p)$ has a Fock decomposition
which is obtained from Eq.~(\ref{Fock}) by the replacement of
the wave functions $\phi_{n }$ by the vertex
functions $\Gamma_{n }$. We can thus cast the eigenstate
equation in the form
\begin{equation}\label{eq3}
{\cal G}(p) = \frac{1}{2\pi}\int
\left[-\tilde{H}^{int}(\omega\tau)\right]\frac{d\tau}{\tau} {\cal G}(p).
\end{equation}
The physical mass $M$ of the compound system is found from the
condition that the eigenvalue is 1. This equation is quite general
and equivalent to the eigenstate equation~(\ref{kt15}). It is
nonperturbative.

\subsection{Spin decomposition of the state vector} \label{spin}
As follows from the angular condition, the spin structure of the
wave functions $\phi_{n}$ is very simple, since its construction
does not require the knowledge of dynamics. It should incorporate
however $\omega$-dependent components. It is convenient to
decompose each wave function $\phi_{n}$ into invariant amplitudes
constructed from all available  particle four-momenta (including
the four-vector $\omega$!) and spin structures (matrices,
bispinors, etc.). In the Yukawa model for instance, we have for
the one- and two-body components~\cite{kms_08}:
\begin{eqnarray}
\label{oneone}
\phi_{1}&= &\psi_1\  \bar{u}(k_1)u(p),\\
\phi_{2}&=&\bar{u}(k_1) \left[ \psi_2  +
\psi'_2\ \frac{M \sla \omega }{\omega \cd p}\right]
u(p), \label{onetwo}
\end{eqnarray}
since no other independent spin structures can be constructed.
Here $\psi_1$, $\psi_2$, and $\psi_2'$ are scalar functions
determined by dynamics. For a spin $1/2$ physical fermion composed
from a constituent spin $1/2$ fermion coupled to scalar bosons,
the number of invariant amplitudes for the two-body Fock component
coincides with the number of independent amplitudes of the
reaction $\mbox{spin}\ 1/2 + \mbox{scalar} \to \mbox{spin} \ 1/2 +
\mbox{scalar}$, which is $(2 \times 2)/2=2$, due to parity
conservation. Similar expansions can be done for the three-body
component of the same system or for Fock components in QED, as we
shall see in Sec.~\ref{appli}.

\section{Fock sector dependent renormalization scheme} \label{FSDR}
In the standard renormalization theory, the bare
parameters\footnote{The term "bare parameters" means here the
whole set of parameters entering into the interaction Hamiltonian,
e.g. the bare coupling constants and the mass counterterms.} are
determined by fixing some physical quantities like the particle
masses and the physical coupling constants.
The bare parameters are thus expressed
thro\-ugh the physical ones. This identification implies in fact
the following two important questions which are usually never
clarified in LFD calculations, but are at the heart of our scheme.

{\it (i)} In order to express the bare
parameters through the physical ones, and vice-versa, one should
be able to calculate observables or, in other words, physical
amplitudes. In LFD, any physical amplitude is represented as a sum
of partial contributions, each depending on the light-front plane
orientation.  Since an observable quantity can not depend on the
latter, this spurious dependence must cancel in the whole sum, as
already mentioned. Such a situation indeed takes place, for
instance, in perturbation theory, provided the regularization of
divergencies in LFD amplitudes is done in a rotationally invariant
way~\cite{kms_07}. In nonperturbative LFD calculations, which are
always approximate, the dependence on the light-front plane
orientation may survive even in calculated physical amplitudes.
For this reason, the identification of such amplitudes with
observable quantities becomes ambiguous and expressing the
calculated amplitudes through the physical parameters turns into a
non-trivial problem.

{\it (ii)} The explicit form of the relationship between the bare
and physical parameters depends on the approximation which is
made. This is a trivial statement in perturbation theory  where
the order of approximation is distinctly determined by the power
of the coupling constant. In our nonperturbative approach based on
the truncated Fock decomposition of the state vector, an analogous
parameter (the power of the coupling constant) is
absent. At the same time, to make calculations compatible with the
order of truncation, one has to trace somehow the level of
approximation. This implies that, on general grounds, the bare
parameters should depend on the Fock sector in which they are
considered.  Moreover, this dependence must be such that all
divergent contributions are cancelled, as already mentioned in
Sec.~\ref{deltam}. How this should be done is the crucial point of
the nonperturbative formulation of field theory on the
light-front.

The use of our FSDR scheme in CLFD is a unique opportunity to
answer both questions. For clarity, we shall take, as a
background, a model of interacting fer\-mi\-ons and bosons like
the Yukawa model or QED, with the aim to calculate the physical
fermion state vector. The basis of Fock space is formed by a set
of Fock sectors, each containing one constituent fermion and a
certain number of bosons (0, 1, 2,...). The truncation is made by
retaining only those Fock sectors where the maximal total number
of constituent particles does not exceed $N$. The consideration of
antiparticle degrees of freedom will be discussed in the simple
case of a scalar model in Sec.~\ref{scalar}. We will assume that
the interaction Hamiltonian $H^{int}(x)$ is constructed through
the bare fields satisfying the free Dirac or Klein-Gordon
equations with the corresponding {\em physical} masses, while the
mass renormalization is performed by introducing, into
$H^{int}(x)$, appropriate mass counterterms.  We emphasize that in
spite of that we consider hereafter several particular forms of
interaction, our scheme is applicable to physical systems with
arbitrary interaction admitting a Fock decomposition of the state
vector.

\subsection{Mass counterterm}
\begin{figure}[btph]
\begin{center}
\includegraphics[width=15pc]{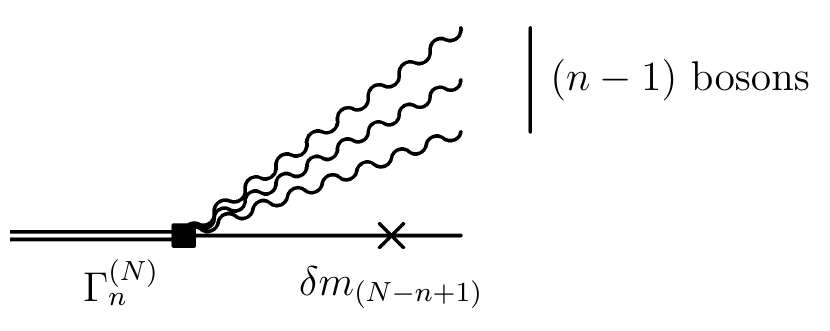}
\caption{Typical insertion of the mass counterterm.}
\label{gammadeltan}
\end{center}
\end{figure}

The simple example of the renormalization of the fermion self-energy
within the two-body Fock space truncation, presented in
Sec.~\ref{deltam}, can serve as a guideline to set up our general
rule. In this example, the mass counterterm should be labelled
with a subscript and denoted by $\delta m_{2}$, in order to
indicate that it is introduced to cancel, at $p\!\!\! /=m$, where
$m$ is the constituent fermion mass, the self-energy contribution
which belongs to the two-body Fock sector. Let us denote by
$\delta m_{l}$ the mass counterterm in the most general case.
Since we truncate our Fock space to order $N$, one should make
sure that, at any light-front time, the total number of
particles is at most $N$. Our first rule is thus:
\begin{itemize}
\item {\it in any amplitude where the mass counterterm  $\delta
m_{l}$ appears, the value of $l$ is such that the total number of
bosons in flight plus $l$ equals the maximal number of the Fock
sectors considered in the calculation, {\it i.e.} $N$.}
\end{itemize}
For instance, in the typical contribution indicated in
Fig.~\ref{gammadeltan},  the mass counterterm  is
 $\delta m_{(N-n+1)}$.
For the mass counterterm of the lowest order, we simply have
\begin{equation}
\label{del12_1} \delta m_{1}=0,
\end{equation}
since the fermion mass is not renormalized at all if the fermion
can not fluctuate in more than one particle!

\subsection{Bare coupling constant}
\label{BCCg}
The general strategy
we developed above for the
calculation of the mass counterterm should be also
applied to the calculation of the bare coupling constant, with however a
bit of caution, since this one may enter in two
different types of contributions.

The first one appears in the calculation of the state vector itself,
 when Eq.~(\ref{eq3}) is solved.
In that case, any boson-fermion coupling constant is associated
with the  emission or the absorption of a boson which participates
in the particle counting, in accordance with the rules detailed
above, since it is a part of the state vector.

The second one appears in the calculation of  the boson-fermion scattering
amplitude or of the boson-fermion three-point Green's function (3PGF)  like the
electromagnetic form factor. Since the
external boson is an (asymptotic) free field rather than a part
of the state vector, the particle counting rule advocated above
should therefore not include the external boson line.

Following the same reasoning developed above for the calculation
of mass counterterms, we can formulate the following general rule
for the calculation of the bare coupling constants
\begin{itemize}
\item {\it in any amplitude which couples constituents inside the state
vector one should attach to each vertex  the internal bare coupling constant
 $g_{0l}$. The value of $l$ is such
that the total number of bosons in flight before (after) the
vertex - if the latter corresponds to the boson emission
(absorption) - plus $l$ equals the maximal number of the Fock
sectors considered in the calculation, {it i.e.} $N$.}
\end{itemize}
The calculation of external bare coupling constants proceeds in
the same spirit, with the final rule:
\begin{itemize}
\item {\it in any amplitude which couples constituents of the
state vector with an external field, one should attach to the
vertex involving this external field the external bare coupling
constant $\bar{g}_{0l}$. The value of $l$ is such that, at the light-front
time corresponding to the vertex, the total number of internal
bosons in flight (those emitted and absorbed by particles entering
the state vector) plus $l$ equals the maximal number of the Fock
sectors considered in the calculation, i.e. $N$.}
\end{itemize}
The lowest order bare coupling constants are
\begin{equation}
g_{01}=0\ ,\ \ \ \bar{g}_{01}=g.\label{g01}
\end{equation}
The first one is trivial, because no fermion-boson interaction is
allowed in the one-body Fock space truncation. The second one
reflects the fact that the external bare coupling constant, in the
same approximation, is not renormalized at all, since a single
fermion can not be "dressed".

\begin{figure}[btph]
\begin{center}
\includegraphics[width=19pc]{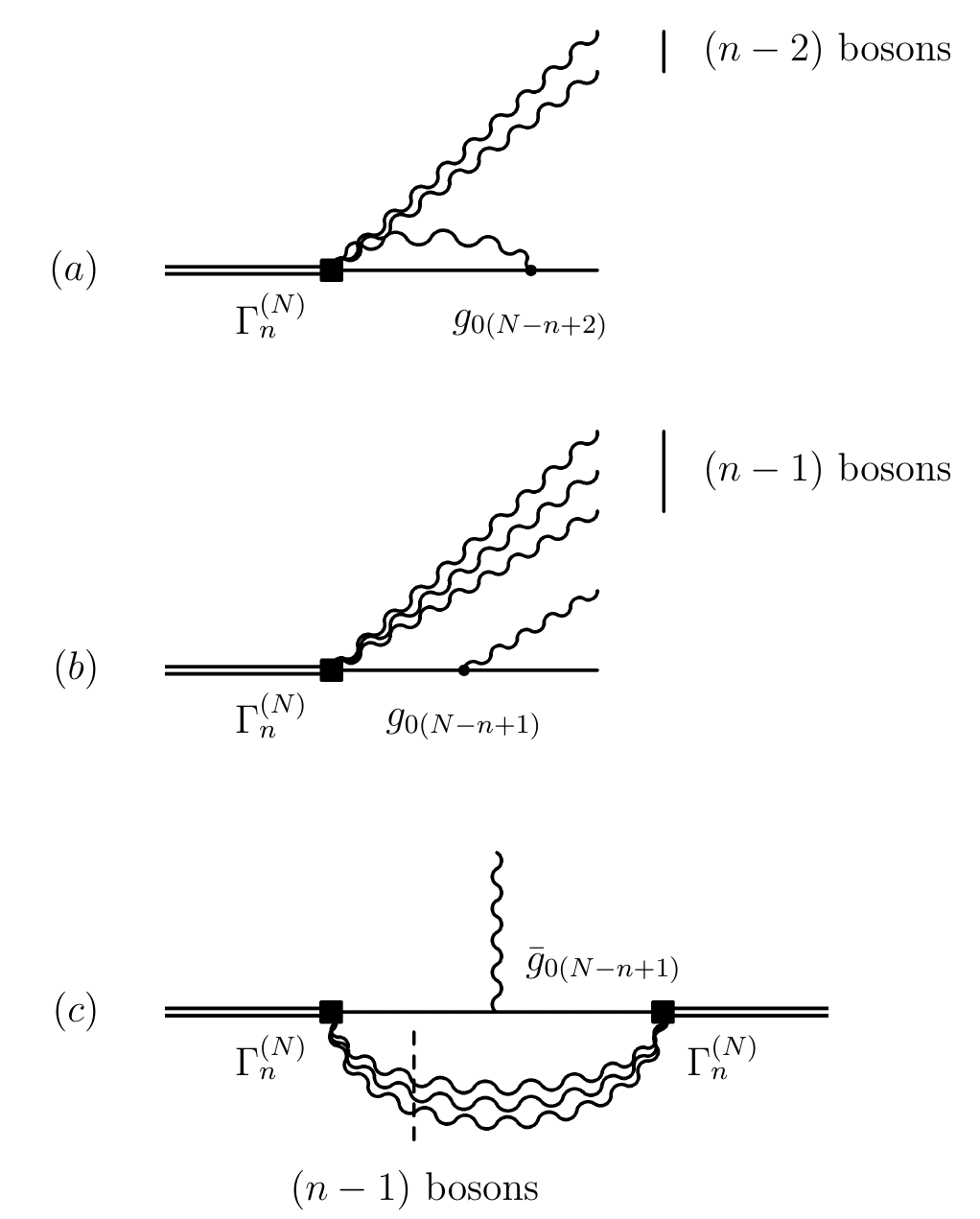}
\caption{Typical contributions to the fermion state vector for the
absorption (a) and the emission (b) of an internal boson, and to
the fermion-boson 3PGF (c). \label{gammag0n}}
\end{center}
\end{figure}

Some illustrations of the rules concerning the internal and
external bare coupling constants are given in Fig.~\ref{gammag0n}.

Though we relied on the fermion-boson model when considering the
above FSDR procedure, the latter can be easily extended to other
systems with additional counterterms and bare parameters.

\subsection{Renormalization conditions and wave function renormalization} \label{renorcond}
Once proper bare coupling constants and mass counterterms  have
been identified, one should fix them from a set of renormalization
conditions. In perturbation theory, there are three types of
quantities to be determined: the mass counterterms, the bare
coupling constants, and the norms of the fermion and boson fields.
Usually, the on-mass-shell renormalization is applied, with the
following conditions. For each field, the mass counterterm is
fixed from the requirement that the corresponding two-point
Green's function has a pole at $p^2=m^2$, where $m$ is the
physical mass of the particle. The field normalization is fixed
from the condition that the residue of the two-point Green's
function at the pole is $1$. The bare coupling constant is
determined by requiring that the on-mass-shell 3PGF is given by
the product of the physical coupling constant and the elementary
vertex.

The renormalization conditions in LFD are of slightly different
form, although they rely on the same grounds. The fermion mass
counterterm is fixed from the eigenvalue equation~(\ref{kt15}) by
demanding that the mass, $M$, of the physical
state is identical to the mass of the constituent fermion which we
called $m$. Solving a similar equation for the physical boson
state vector allows us to find the boson mass counterterm from
quite analogous requirements. The determination of the internal
bare coupling constant needs some care. It is found, as in
perturbation theory, by relating the on-energy-shell two-body
vertex function $\Gamma_2$ to the physical coupling constant $g$.
As follows from the momentum conservation law, taking $\Gamma_2$
on the energy shell is equivalent to setting $\tau_2=0$, or
$s_2=(k_1+k_2)^2=M^2$.

In order to fix the relationship between $\Gamma_2$ and $g$ one
needs to take into account the renormalization factors coming from
radiative corrections to all legs of the two-body vertex
function~\cite{hb_99}. These factors do also depend on the order
of the Fock space truncation, as detailed in~\cite{yukawa}. In the
case of a fermion coupled to bosons, without polarization effects
generated by antifermion contributions, this relationship reads
\begin{equation} \label{gamma2r}
\Gamma_2^{(N)}(s_2=M^2)=g \sqrt{I_1^{(N-1)}},
\end{equation}
where $I_1$ is the one-body contribution to the norm of the state
vector, as given in Eq.~(\ref{Inor}), calculated for the Fock
space truncation of order $N-1$.

Eq.~(\ref{gamma2r}) admits simple physical interpretation. Each
leg of the on-energy-shell two-body vertex function contributes
for an individual factor $\sqrt{Z}$ to the physical coupling
constant, where $Z$ is the field strength normalization
factor~\cite{ps}. The physical fermion state is normalized to $1$,
so that its factor $Z$ equals $1$. The constituent boson line is not
renormalized - since we do not consider antifermions - so that for the
bosonic line we also have $Z_b=1$. If we did not neglect the
antifermionic degrees of freedom, it would contribute by a
non-unity factor $\sqrt{Z_b}$. Finally, we have shown
in~\cite{yukawa} that the field strength normalization factor of
the constituent fermion is just the weight of the one-body
component in the norm of the physical state, i.e. $Z_f=I_1$.
According to our FSDR scheme, the normalization factor of the
constituent fermion should correspond to the Fock space truncation
of order $N-1$, since, by definition,  the two-body vertex function contains one
extra boson in flight in the final state.

The condition~(\ref{gamma2r}) has two important consequences. The
first one is that the two-body vertex function at $s_2=M^2$ should
be independent of the four-vector $\omega$ which determines the orientation of the
light-front plane. With the spin decomposition~(\ref{onetwo}),
this implies that the component $\psi_2^\prime$ at $s_2=M^2$
should be identically zero. While this property is automatically
verified in the case of the two-body Fock space truncation - if
using a regularization scheme which does not violate rotational
invariance - this is not guaranteed for calculations within higher
order truncations.

Indeed, nothing prevents $\Gamma_2$ to be $\omega$-dependent,
since it is an off-shell object, but this dependence must
completely disappear on the energy shell, i.e. for $s_2=M^2$. It
would be so if no Fock space truncation occurs. The latter, in
approximations higher than the two-body one (i.~e. for
$N=3,\,4,\ldots$), may cause some $\omega$-dependence of
$\Gamma_2$ even on the energy shell, which immediately makes the
general renormalization condition~(\ref{gamma2r}) ambiguous. If
so, one has to insert new counterterms into the light-front interaction
Hamiltonian, which explicitly depend on $\omega$
and cancel the $\omega$-dependence of $\Gamma_2(s_2=M^2)$. Note
that the explicit covariance of CLFD allows to separate the terms
which depend on the light-front plane orientation (i.e. on
$\omega$) from other contributions and establish the structure of
these counterterms. This is not possible in ordinary LFD.

We should thus enforce the condition~(\ref{gamma2r}) by
introducing, into the interaction Hamiltonian, appropriate
$\omega$ dependent counterterms. For instance, in the Yukawa model
without antifermion contributions and within the
three-body Fock space truncation (see below, Sec.~\ref{Yukawa3}),
we need one additional counterterm of the
form~\cite{kms_08,yukawa}
\begin{equation}
\label{Zomega} \delta {H}^{int}_\omega = -Z_\omega \bar \psi
\frac{m \sla \omega}{i \omega \cd
\partial} \psi\varphi,
\end{equation}
where $Z_\omega$ is a constant adjusted to make Eq.~(\ref{gamma2r})
true,
$\psi (\varphi)$ is the fermion (boson) field, and $1/i(\omega\cd
\partial)$ is the reversal derivative operator, fully analogous to
the operator $1/i\partial^+$ in ordinary light-front dynamics.

The second consequence of Eq.~(\ref{gamma2r}) is that the two-body
vertex function at $s_2=M^2$ should be a constant. This is a
non-trivial requirement, since $\Gamma_2$, as well as the
components $\psi_2$ and $\psi_2^\prime$ in Eq.~(\ref{onetwo}), do
depend on two invariant kinematical variables which are usually
chosen as the longitudinal constituent momentum fraction, $x$, and
the square of its transverse momentum, ${\bf k}_\perp^2$. The
on-mass-shell condition
\begin{equation}
\label{s2}
s_2=\frac{{\bf k}_{\perp}^2+\mu^2}{x}+\frac{{\bf k}_{\perp}^2+m^2}{1-x}=M^2,
\end{equation}
where $\mu$ is the constituent boson mass, fixes one variable
(say, ${\bf k}_{\perp}^2$) only, while the second variable ($x$) remains
arbitrary. Hence, under the condition~(\ref{s2}), if we assume
${\bf k}_{\perp}^2$ fixed, $\Gamma_2$ should be independent of $x$.
Again, this property is verified in the two-body Fock space
truncation, since in this approximation our equations are
equivalent to perturbation theory of order $g^2$. It is not
guaranteed for higher order calculations. We shall come back to
this point in Sec.~\ref{appli}. In practice, we shall fix
$\Gamma_2(s_2=M^2)$ at some preset value $x^*$ and verify that the
physical observables are not sensitive to the choice of $x^*$.

To summarize, we can thus list the normalization conditions in CLFD,
for a calculation done in a Fock
space truncation of order $N$, without considering fermion-antifermion
polarization corrections:

{\it
\begin{itemize}
\item The mass counterterm $\delta m_{N}$ is fixed by solving the
eigenstate equation~(\ref{eq3}) in the limit $M\to m$. \item The
state vector is normalized according to the standard
condition~(\ref{Inor}). \item The internal bare coupling constant
$g_{0N}$ is fixed from the condition that the $\omega$-independent
part of the two-body vertex function at $s_2=m^2$ and at a fixed
value of $x$, denoted by $x^*$, is given by the right-hand side of
Eq.~(\ref{gamma2r}). \item The external bare coupling constant
$\bar{g}_{0N}$ is fixed from the condition that the
$\omega$-independent part of the on-energy-shell 3PGF is
proportional to the elementary vertex, with the proportionality
coefficient being the physical coupling constant. \item The
$\omega$-dependent counterterms in the Hamiltonian are fixed by
the conditions that the $\omega$-dependent parts of the two-body
vertex function and the 3PGF, both taken on the energy shell, turn
to zero. \item The values of all bare parameters and
counterterms for $l\leq N$ are determined from successive
calculations within the $1,2,\dots N$ Fock space truncations.
\end{itemize}
}

\section{Applications} \label{appli}
The explicit solution of the eigenvalue equation~(\ref{eq3})
requires to define first a regularization scheme, since loop
contributions are {\em a priori} divergent. We shall use in the
following applications the Pauli-Villars (PV) regularization
scheme, first applied to LFD in~\cite{BHMc}. It has the nice feature
of being rotationally invariant~\cite{kms_07}, and it can be
implemented rather easily in calculations within the two- and
three-body Fock space truncations~\cite{kms_08}. Besides that, in
this regularization scheme, all contact interactions inherent to
LFD are absent. It however necessitates to extend Fock space in
order to embrace PV fermions and PV bosons on equal grounds with
the physical particles.

\subsection{Self-energy of a fermion in the Yukawa model}
Before discussing the calculation of the properties of compound
systems, it is instructive to look at the structure of the fermion
self-energy in the simple Yukawa model in CLFD. The results are
very similar to those for QED~\cite{kms_07}.

Since our formalism is explicitly covariant, we can write down
immediately the general structure of the self-energy of a fermion
with the off-shell four-momentum $k$ ($k^2\neq m^2$). It writes
\begin{equation}
\label{Sigp} \Sigma({\sla k}) = {\cal A}+ {\cal B} \frac{\sla
k}{m} + {\cal C}\; \frac{m{\sla \omega}}{\omega\cd k}+ {\cal
C}_1\sigma,
\end{equation}
where $\sigma=({\sla k}{\sla \omega}-{\sla \omega}{\sla
k})/4(\omega\cd k)$. The coefficients in this expansion depend on
$k ^2$ only. They are given by
\begin{eqnarray}
\label{ABC1} {\cal A}(k^2)&=&\frac{1}{4}
\mbox{Tr}\left[\Sigma({\sla k})\right],\\
\label{ABC2} {\cal B}(k^2)&=&\frac{m}{4 \omega\cd k}
\mbox{Tr}\left[\Sigma({\sla k}){\sla
\omega}\right],\\
\label{ABC3} {\cal
C}(k^2)&=&\frac{1}{4m}\mbox{Tr}\left[\Sigma({\sla k})\left({\sla
k}-\sla{\omega}\frac{k^2}{\omega\cd k}\right)\right],\\
\label{C} {\cal C}_1(k^2)& =& \mbox{Tr}\left[\Sigma({\sla
k})\sigma\right].
\end{eqnarray}
In the two-body approximation, when $\Sigma({\sla k})$ is entirely
given by the loop diagram shown in Fig.~\ref{self}, the
coefficient ${\cal C}_1$ is identically zero. The coefficient
${\cal C}$ should also be zero since the two-point Green's
function should be equivalent to the one calculated in the
four-dimensional Feynman approach, and is therefore independent of
$\omega$ provided one uses a rotationally invariant
regularization~\cite{kms_07}. Note that the coefficient ${\cal C}$
is not {\em a priori} chiral invariant in the sense that if the
mass of the constituent fermion goes to zero, ${\cal C}$ does not
vanish, in contrast to ${\cal A}$ and ${\cal B}$, as it should.
Using the PV regularization scheme, ${\cal A}$ and ${\cal B}$
depend logarithmically on the PV boson mass. Without
regularization, the coefficient $C$ diverges quadratically at high
momenta. It is however identical to zero when the PV regularization
scheme is used with one PV fermion and one PV boson only.

\subsection{QED in the two-body approximation} \label{QED2}
\begin{figure}[ht!]
\begin{center}
\includegraphics[width=20.5pc]{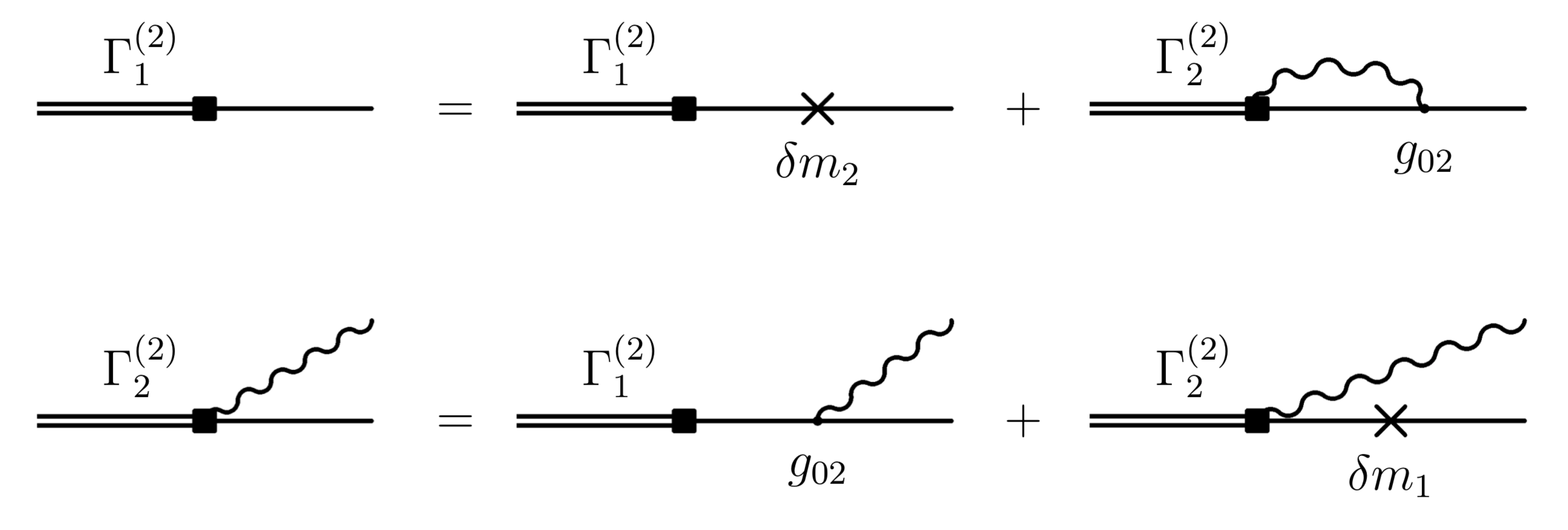}\caption{System of equations for the vertex
functions in the two-body approximation.}\label{fig1}
\end{center}
\end{figure}

\begin{figure}[btph]
\begin{center}
\includegraphics[width=18pc]{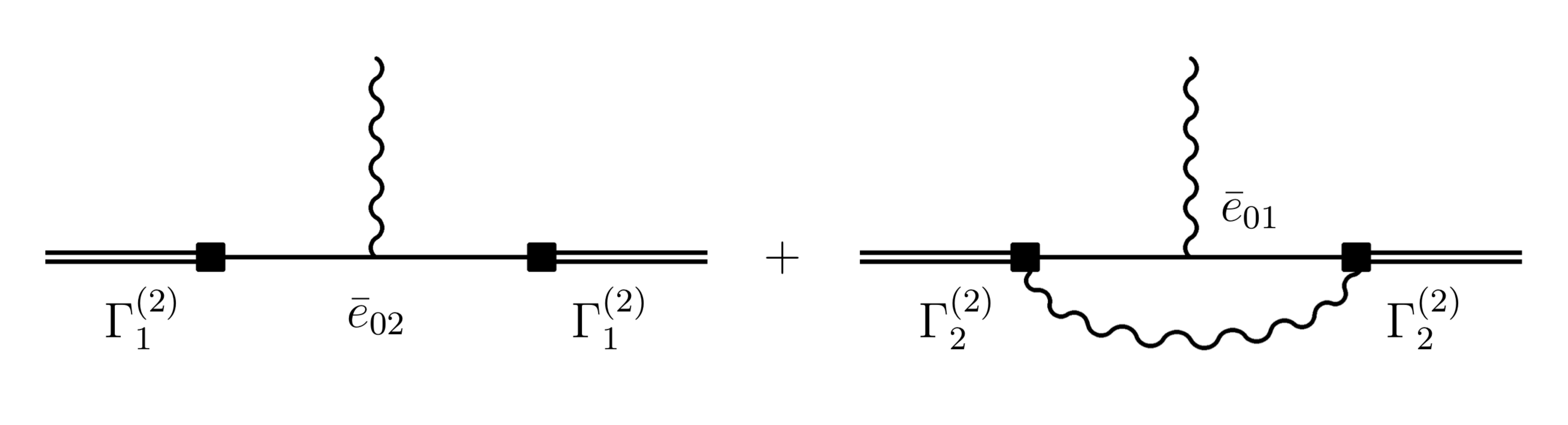}
\caption{Fermion-boson 3PGF in the two-body approximation.}
\label{ff2}
\end{center}
\end{figure}

This simple case provides a good starting point to understand how
our general framework should be applied in practice. The
eigenstate equation one has to solve is shown graphically in
Fig.~\ref{fig1}. This equation is written in accordance with the
prescriptions of our FSDR scheme. Note the use of the Fock sector
dependent mass counterterms  and the bare coupling constants. The
two-body vertex function writes in that case
\begin{equation}
\bar{u}(k_1)[\Gamma_2^{\mu}e_{\mu}^{\lambda}(k_2)]u(p),
\end{equation}
where $e_{\mu}^{\lambda}$ is the polarization vector of the
photon. To simplify notations, we omit the superscript ``(2)'' at
$\Gamma_2^\mu$. We shall use in the following the Feynman gauge.
The number of independent invariant amplitudes for this vertex
function coincides with that for the reaction $1/2+0\to 1/2+1$.
However, one should take into account that in the Feynman gauge
the vector boson wave function has {\em four} independent
components. So, the total number of invariant amplitudes is
$(2\times 2\times 4)/2=8$. We choose the following set of
invariant amplitudes~\cite{kms_04}:
\begin{eqnarray}
\Gamma_2^\mu&=&
b_1\gamma^{\mu}+b_2\frac{m\omega^{\mu}}{\omega\cd p} +
b_3\frac{m\sla{\omega}\gamma^{\mu}}{\omega\cd p}+
b_4\frac{m^2\sla{\omega}\omega^{\mu}}{(\omega\cd p)^2}\nonumber \\
&&+ b_5\frac{p^{\mu}}{m}+b_6\frac{k_1^{\mu}}{m}+
b_7\frac{m\sla{\omega}p^{\mu}}{\omega\cd p}
+b_8\frac{m\sla{\omega}k_1^{\mu}}{\omega\cd p}.
\end{eqnarray}
In the two-body approximation, and using the PV regularization
scheme, we find
\begin{eqnarray}
b_1&=&2\,e_{02} \, m \, \psi_1, \\
b_{2-8}&=&0,
\end{eqnarray}
where $\psi_1$ is defined in Eq.~(\ref{oneone}). These components
refer to the physical ones. The ones associated with PV bosons
and/or fermions can be found easily~\cite{kms_08}.
In this approximation, the vertex function
$\Gamma_2^{\mu}$ is a constant matrix proportional to
$\gamma^{\mu}$. With these results, one can calculate the norms of the Fock
sectors entering into the normalization
condition~(\ref{Inor}):
\begin{eqnarray}
I_1&=&4m^2 \psi_1^2,\\
I_2&=&4m^2 \psi_1^2\,e_{02}^2\,J_2,
\end{eqnarray}
where the expression for $J_2$ can be found in~\cite{kms_04}. It
depends logarithmically on the mass of the PV boson used to
regularize the loop integral. The normalization condition thus
fixes $\psi_1$:
\begin{equation} \label{psi1}
4m^2\psi_1^2=\frac{1}{1+e_{02}^2J_2}.
\end{equation}
The renormalization condition~(\ref{gamma2r}) enables us to
calculate $e_{02}$ as a function of the physical coupling constant
denoted by $e$. This condition, for $N=2$, writes simply
\begin{equation}
\Gamma_2^{\mu}=e \gamma^\mu\equiv 2 e_{02}m \psi_1 \gamma^\mu,
\end{equation}
since $I_1^{(1)}=1$. With Eq.~(\ref{psi1}) we get
\begin{equation}
e_{02}^2=\frac{e^2}{1-e^2 J_2}
\end{equation}
and
\begin{eqnarray}
I_1&=&1-e^2J_2, \\
I_2&=&e^2J_2.
\end{eqnarray}
With all these quantities, one can easily calculate the
electromagnetic form factors given by the diagrams shown in
Fig.~\ref{ff2}. Note that since the state vector is normalized to
$1$, we find for the external bare coupling constant:
\begin{equation}
\bar e_{01}=\bar e_{02}=e.
\end{equation}
At zero momentum transfer, the Pauli form factor gives the
anomalous magnetic moment of the electron. We get~\cite{kms_08}
\begin{equation} \label{schwinger}
F_2(0)=\frac{\alpha}{2\pi}.
\end{equation}
This expression exactly coincides with the well-known Schwinger
term. One may wonder why our nonperturbative approach exactly
recovers the perturbative result~\cite{ch_09}. The reason is
simple. For the two-body Fock space truncation we consider here,
the irreducible contributions to the two-point Green's function or
to the electromagnetic form factors are just identical to the
corresponding ones in the second order of perturbation theory. And
their re-summation to all orders of the coupling constant just
defines the physical mass of the electron, which is also the same,
by construction, in both approaches.

Our result~(\ref{schwinger}) is also independent of the PV fermion
and boson masses, when they tend to infinity. This is of course
necessary in order to preserve the scale invariance of physical
observables. Note that at large enough PV masses, the one-body
part $I_1$ of the normalization condition is negative, while the
two-body part $I_2$ exceeds $1$. This implies also that $e_{02}^2$
is negative. These features are just an artefact of the
regularization scheme which is used. They show that indeed $I_1$
and $I_2$, and, more generally, any $I_n$ in Eq.~(\ref{Inor}), as
well as the bare coupling constants, are not physical observables
and can therefore be scale dependent.

\subsection{Scalar system in the three-body approximation:
the role of antiparticles} \label{scalar}

Finding the state vector for the three-body Fock space truncation
is the first non-trivial nonperturbative calculation. We begin our
consideration of the three-body approximation with the study of a
toy model: a heavy scalar boson $B$ with mass $m$, interacting
with light scalar bosons $b$ with mass $\mu$. We will calculate
the state vector of the heavy boson and represent the former as a
sum of the following Fock sectors, in symbolic form:
\begin{equation}
\label{eq4-1}
\phi(p)=|B\rangle+|Bb\rangle+|Bbb\rangle+|BB\bar{B}\rangle,
\end{equation}
where $\bar{B}$ denotes the antiparticle. The interaction
Hamiltonian is
\begin{equation}
\label{eq4-2} H^{int}(x)=-g_0B^2(x)b(x)-\delta m^2B^2(x)-\delta
\mu^2 b^2(x),
\end{equation}
where $\delta m^2$ and $\delta \mu^2$ are the mass counterterms
which, in this model, have the dimension of mass squared. Since
the physical coupling constant $g$ has the dimension of mass, it
is convenient to define a dimensionless coupling constant
$\alpha=g^2/16\pi m^2$.

\begin{figure}[btph]
\begin{center}
\includegraphics[width=19pc]{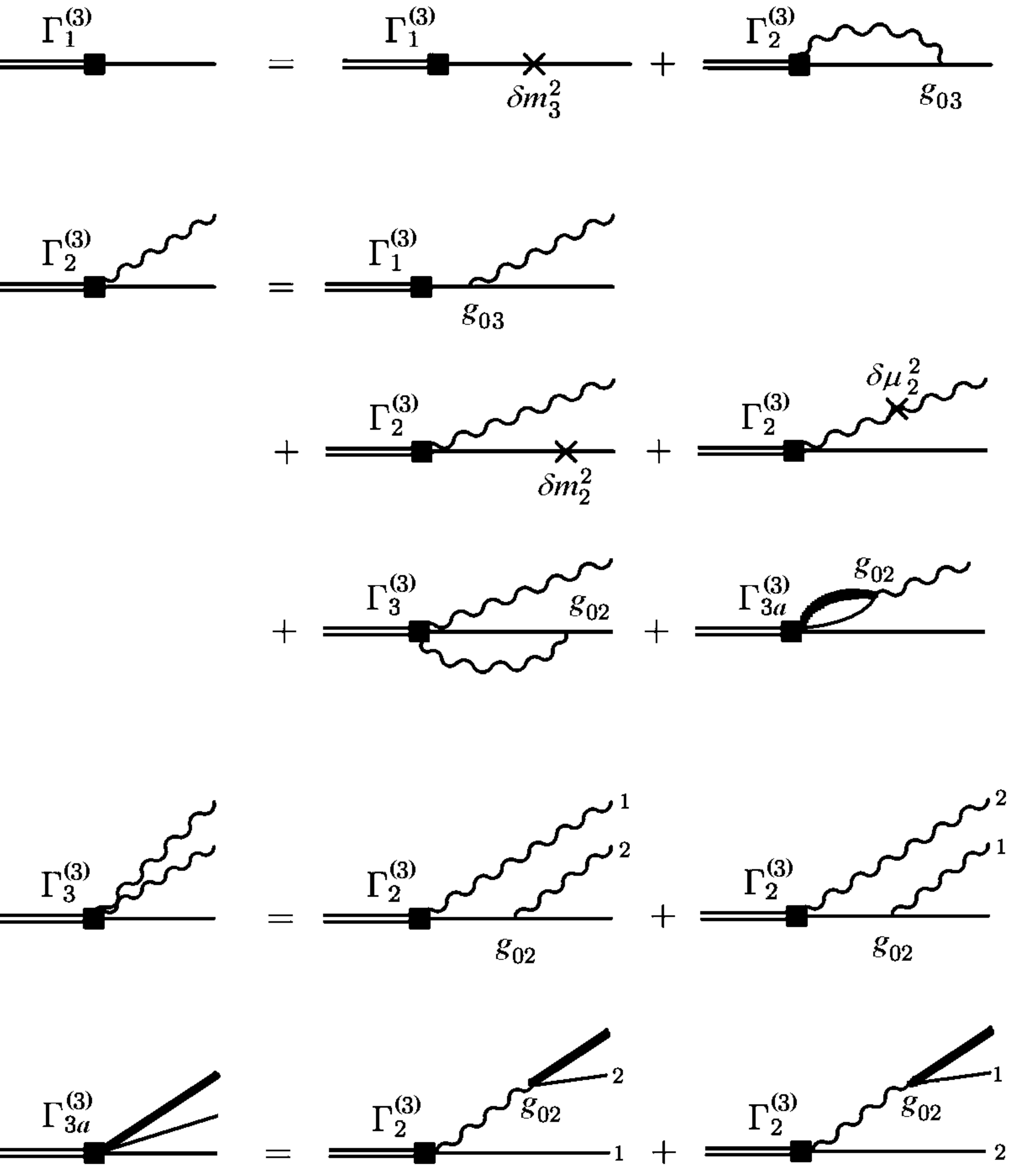}
\caption{System of equations for the vertex functions of the
scalar system in the three-body approximation. The thin and thick
ordinary lines stand for heavy bosons and antibosons,
respectively. The wavy lines denote light bosons.} \label{syseq3}
\end{center}
\end{figure}

The system of eigenstate equations for the vertex functions is
represented graphically in Fig.~\ref{syseq3}. The two three-body
vertices $\Gamma_3^{(3)}$ and $\Gamma_{3a}^{(3)}$ are trivially expressed
through the two-body vertex $\Gamma_2^{(3)}$. Due to this fact, we can
obtain a closed equation involving $\Gamma_2^{(3)}$ only.

\begin{figure}[btph]
\begin{center}
\includegraphics[width=12pc]{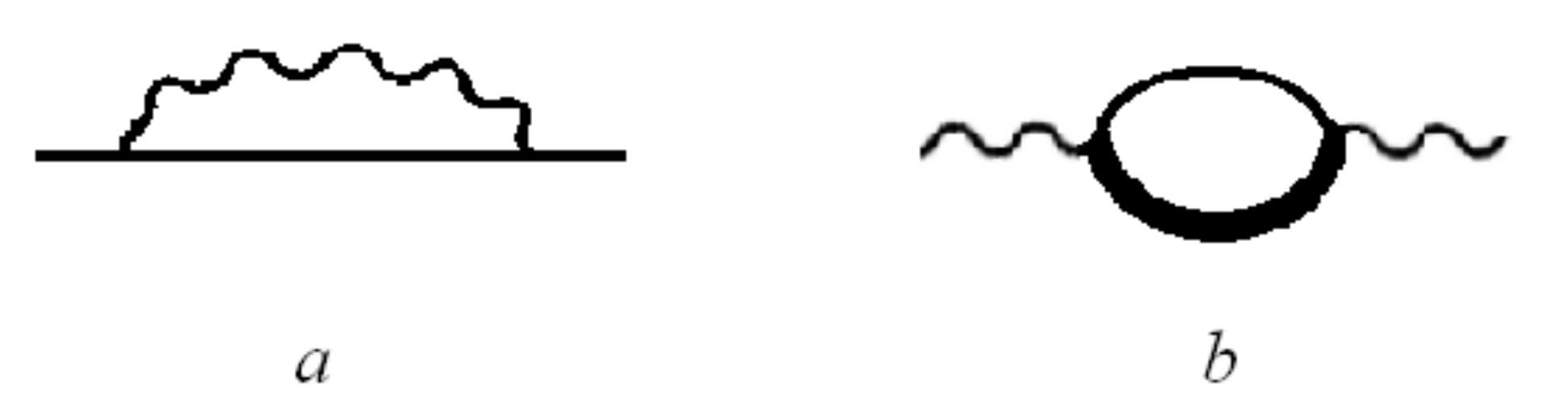}
\caption{Divergent contributions in the scalar model: the heavy
(a) and light (b) boson self energies.} \label{BbSE}
\end{center}
\end{figure}

In this model, we encounter two types of irreducible
divergent contributions: the heavy and light boson self-energies, as shown in Fig.~\ref{BbSE}. Both
of them diverge logarithmically at high momenta; the divergences
are expected to be cancelled by the corresponding mass
counterterms.

The renormalization condition should  now incorporate the light
boson field strength normalization factor. Instead of
Eq.~(\ref{gamma2r}), we have
\begin{equation}
\label{eq4-3} \Gamma_2^{(3)}(s_2=M^2)=g\sqrt{Z_{Bb}},
\end{equation}
where the factor $Z_{Bb}$ stands for the combined contribution of
the heavy and light boson strength normalization factors,
calculated according to the FSDR prescription, extended to include
antiparticle degrees of freedom~\cite{kms_10}. This factor is
completely finite in a pure scalar system, and is expressed in
terms of derivatives of the heavy and light boson self-energies.

The eigenstate equation for $\Gamma_2^{(3)}$ with the renormalization
condition~(\ref{eq4-3}) has been solved numerically in the limit
$M\to m$ for the following set of parameters: $m=0.95$,
$\mu=0.15$, $\alpha=2$. Note that this value of $\alpha$ is far
from the perturbative domain. For convenience, $\Gamma_2^{(3)}$ is
parameterized as a function of two variables, $s_2$ and $x$. The
physical domain corresponds to $s_2\geq (m+\mu)^2$, $0<x<1$.
The renormalization point is defined by $s_2=m^2$ and $x=x^*$, where we chose
$x^*=\mu/(m+\mu)$.

\begin{figure}[btph]
\begin{center}
\includegraphics[width=20pc]{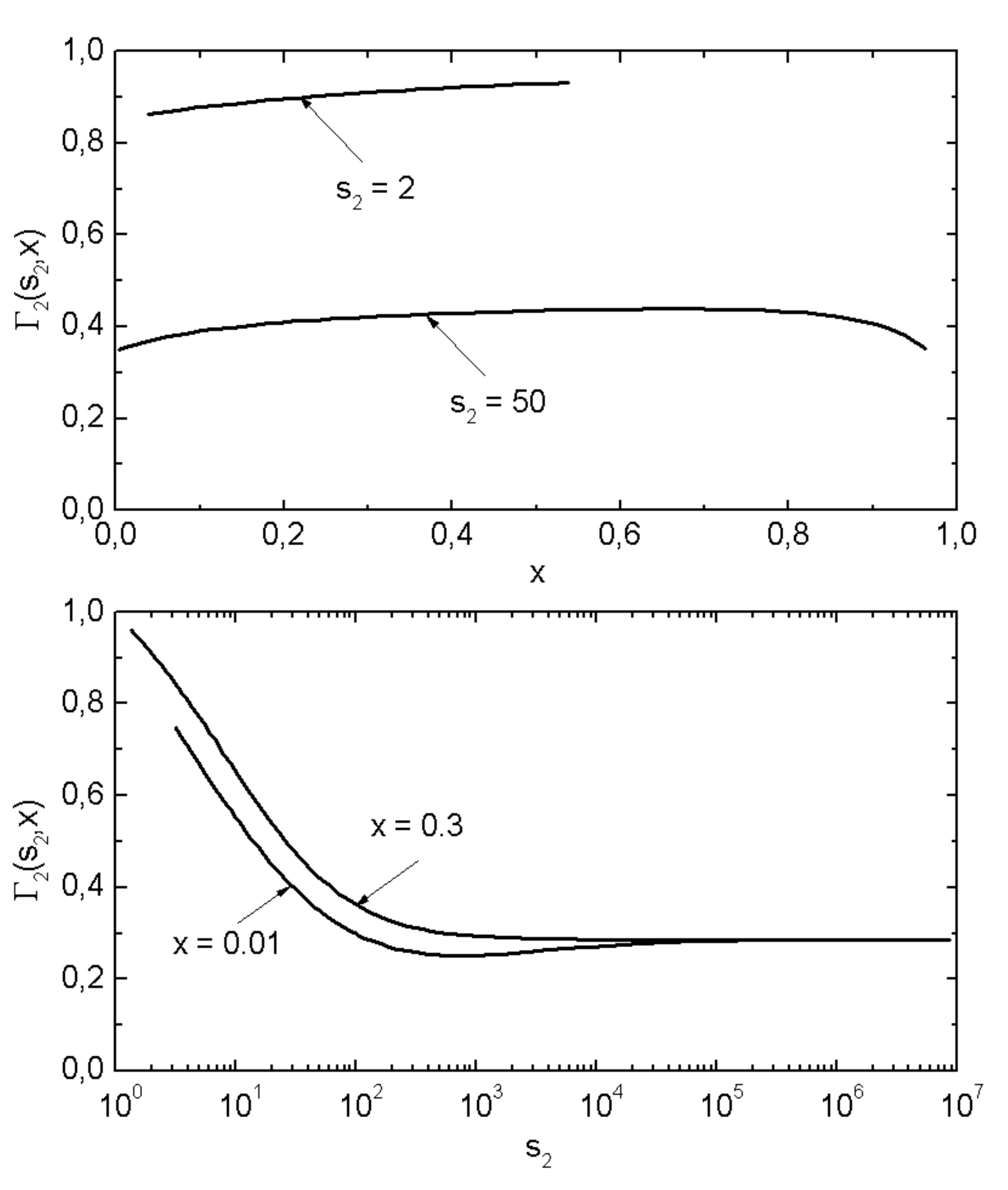}
\caption{Two-body vertex function in the scalar model as a
function of $x$ at fixed $s_2$ (upper plot) and as a function of
$s_2$ at fixed $x$ (lower plot).} \label{G2sx}
\end{center}
\end{figure}

We show in Fig.~\ref{G2sx}  the characteristic dependence of
$\Gamma_2^{(3)}$ on one of its arguments, while the second argument is
fixed. In contrast to the two-body case, where $\Gamma_2^{(2)}$ is a
constant, now it exhibits rather nontrivial behavior, especially
as a function of $s_2$ at fixed $x$.

\begin{figure}[btph]
\begin{center}
\includegraphics[width=20pc]{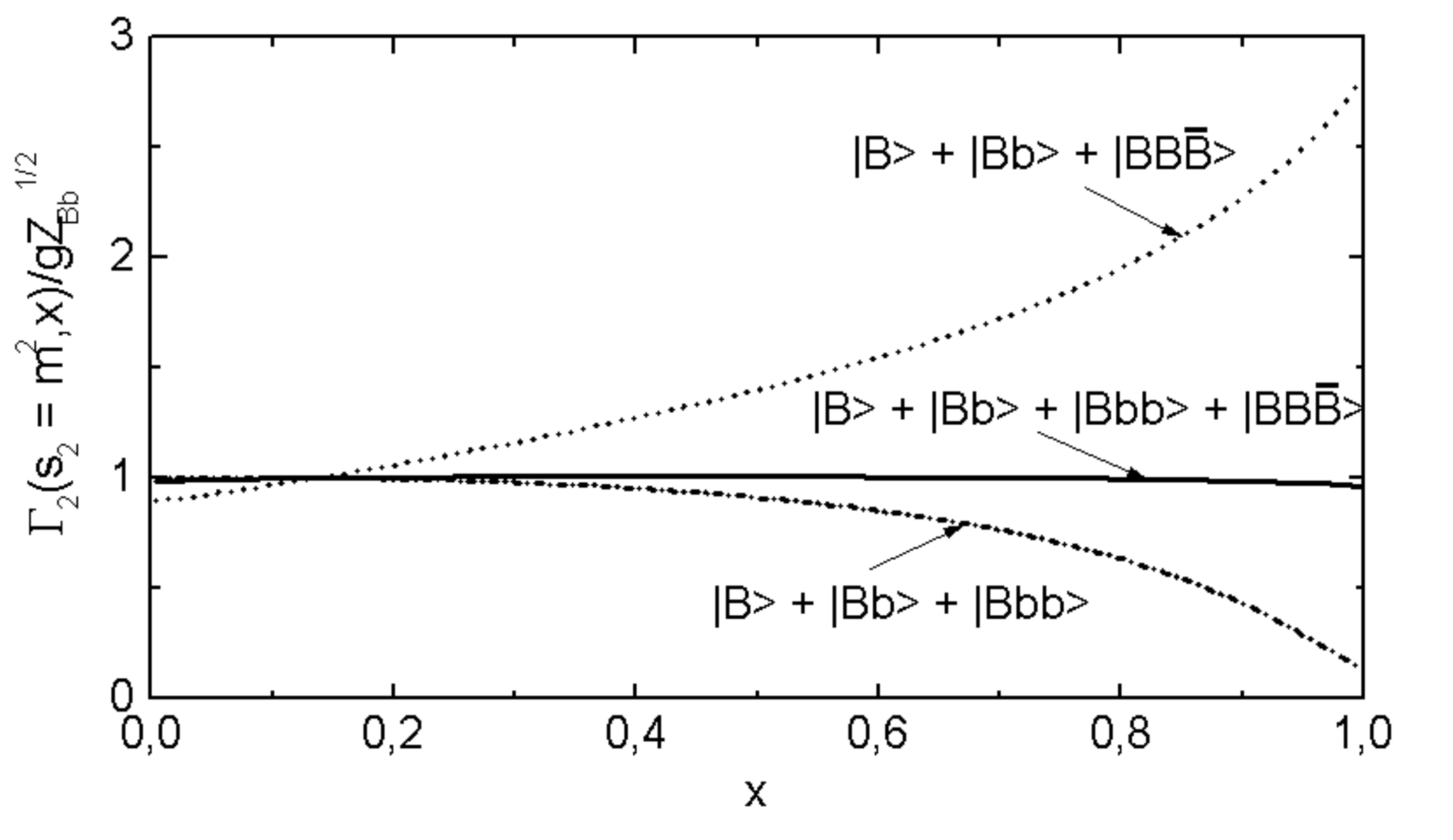}
\caption{Two-body vertex function (in units of $g\sqrt{Z_{Bb}}$)
at $s_2=m^2$ as a function of $x$. The solid line is the
calculation with all the four Fock sectors, Eq.~(\ref{eq4-1}), in
the state vector. The dash-dotted and dotted lines are the same
but without the $|BB\bar{B}\rangle$ or $|Bbb\rangle$ sector,
respectively.} \label{G2x_st}
\end{center}
\end{figure}

To check the consistency of the renormalization
condition~(\ref{eq4-3}),  we show in Fig.~\ref{G2x_st} the
dependence of $\Gamma_2^{(3)}$ on $x$ at fixed $s_2=m^2$. It is
distinctly seen that if the Fock space includes all the four Fock
sectors, Eq.~(\ref{eq4-1}), $\Gamma_2^{(3)}(s_2=m^2,x)$ is almost a
constant, as it ought to be. If one however neglects one of the
three-body sectors, either $|BB\bar{B}\rangle$ or $|Bbb\rangle$,
$\Gamma_2^{(3)}(s_2=m^2,x)$ strongly depends on $x$, which makes the
renormalization condition ambiguous because physical results turn
out to be sensitive to the choice of $x^*$.

\begin{figure}[bth]
\begin{center}
\includegraphics[width=8.5cm]{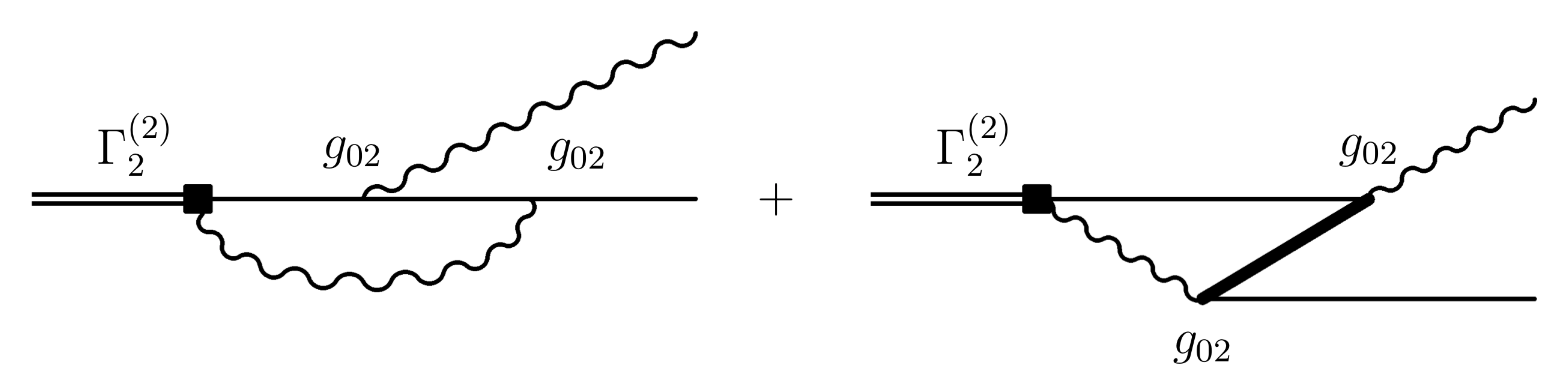}
\end{center}
\caption {Part of the three-body Fock sector contributions to the two-body
vertex.}\label{anti}
\end{figure}

The above result is very non-trivial. Let us consider, for
instance, a part of the three-body contributions to
$\Gamma_2^{(3)}$ calculated in perturbation theory, i.e. in terms
of the two-body vertex function $\Gamma_2^{(2)}$ found in the
lower, two-body, approximation, as shown in Fig.~\ref{anti}. Their
sum (but not each of them!) is exactly $x$-independent at
$s_2=m^2$, since $\Gamma_2^{(2)}$ is a constant. This statement
follows from the fact that the sum of the two contributions  on
the energy shell coincides with the corresponding on-mass-shell
Feynman amplitude, and the latter is a constant. In our
calculation of order $N=3$, the contributions analogous to those
in Fig.~\ref{anti}, but with $\Gamma_2^{(3)}$ instead of
$\Gamma_2^{(2)}$ appear in the eigenstate equation and determine
the $x$-dependence of $\Gamma_2^{(3)}(s_2=m^2)$. From
Fig.~\ref{G2x_st} it is seen that the latter dependence is
surprisingly weak, though $\Gamma_2^{(3)}$ is not a constant, as
Fig.~\ref{G2sx} clearly demonstrates. At the same time, if one
calculates the amplitudes of the diagrams in Fig.~\ref{anti} with
an arbitrary chosen function $\Gamma_2(s_2,x)$ instead of
$\Gamma_2^{(2)}$, their sum would hardly be a constant at
$s_2=m^2$. Such a result may indicate that the set of
contributions we considered in the case of the three-body Fock
space truncation forms an almost consistent set, in the sense that
it meets all consistency requirements. This constitutes a basis of
a well-controlled, nonperturbative Fock expansion of the the state
vector, in the same spirit as an expansion in $g^2$ in
perturbation theory.
\subsection{Yukawa model in the three-body approximation: calculation of the fermion
anomalous magnetic moment} \label{Yukawa3}

\begin{figure}[ht!]
\begin{center}
\includegraphics[width=6cm]{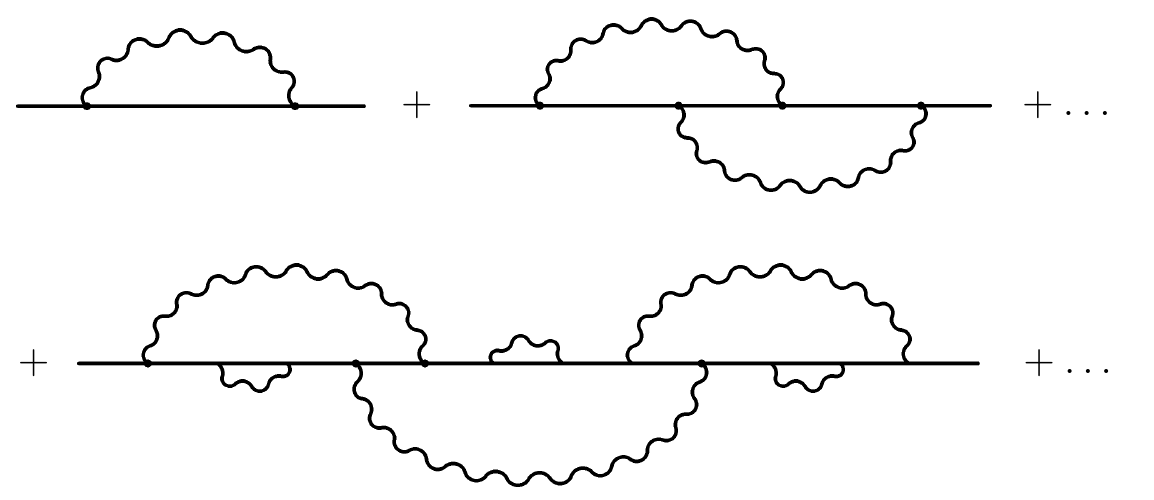}
\end{center}
\caption{Radiative corrections to the self-energy.\label{fig4}}
\end{figure}

The Yukawa model is much closer to reality (e.g. for instance to QED) than the
scalar model discussed in the previous section. Simultaneously, it
is much more involved from the point of view of renormalization,
because of complicated spin structure of the vertex functions and
stronger divergences to be regularized and renormalized. In this
section, we apply our FSDR renormalization scheme to the Yukawa
model within the three-body Fock space truncation, but, in order
to avoid extra complications, without incorporating antifermion
degrees of freedom. Our purpose is to demonstrate the
capabilities of the FSDR scheme in solving a true nonperturbative
problem for particles with spin. For example, in contrast to the
two-body Fock space truncation (in this approximation the Yukawa
model is almost equivalent to QED, as considered in
Sec.~\ref{QED2}), where we have only one irreducible contribution
to the fermion self-energy, we now should consider all the graphs
for the self-energy, which contain one fermion and two bosons in
intermediate states, including overlapping self-energy type
diagrams. The number of such irreducible graphs is infinite. Some
of them are shown in Fig.~\ref{fig4}. The  solution of the
eigenstate equation for the state vector automatically generates
these contributions to all orders of the coupling constant.

We start with the following Fock decomposition for the state vector
\begin{equation}
\label{eq4-5} \phi(p)=|F\rangle+|Fb\rangle+|Fbb\rangle,
\end{equation}
where the symbols $F$ and $b$ denote, respectively, the
constituent fermion and boson with masses $m$ and $\mu$. The
interaction Hamiltonian reads
\begin{equation}
\label{eq4-6} H^{int}(x)=-g_0\bar{\psi}\psi\varphi-\delta
m\bar{\psi}\psi-\delta \mu^2 \varphi^2+\delta {H}^{int}_\omega,
\end{equation}
where $\delta {H}^{int}_\omega$ is defined by Eq.~(\ref{Zomega}).
The regularization is done in a rotationally invariant way by
introducing one PV fermion with  mass $m_1$ and one PV boson
with mass $\mu_1$.

The system of equations for the vertex functions in graphical
representation can be obtained from that for the scalar case,
shown in Fig.~\ref{syseq3}, by changing $\delta m_{2,3}^2\to
\delta m_{2,3}$, $g_{03}\to g_{03}+Z_{\omega}m{\sla
\omega}/\omega\cd p$ and by setting $\delta \mu^2_2\equiv 0$,
$\Gamma_{3a}^{(3)}\equiv 0$, since we do not consider here the dressing of the boson
line. After that, the expressions for the light-front
diagrams are written according to the CLFD graph technique rules,
taking into account that certain lines correspond now to a
particle with spin 1/2.

Expressing $\Gamma_3^{(3)}$ through $\Gamma_2^{(3)}$, as in the
scalar case, we get a closed matrix equation for the two-body
vertex. Its spin structure is written similarly to
Eq.~(\ref{onetwo}):
\begin{equation}
\label{eq4-7} \Gamma_2^{(3)}=\bar{u}(k_1)\left[b_1+b_2\frac{m{\sla
\omega}}{\omega\cd p}\right]u(p),
\end{equation}
where $b_1$ and $b_2$ are scalar functions. For clarity, we
indicate here only the physical components. The PV fermion and
boson components can be calculated as well~\cite{yukawa}.

The three-body component is completely determined by four scalar
functions, like, for instance
\begin{eqnarray}
\label{wf2}
&&\bar{u}(k_{1})\Gamma_3^{(3)}(1,2,3) u(p) \\
 &&=\bar{u}(k_{1})\Bigl(g_1\,S_1+g_2\,S_2+g_3\,S_3+g_4\,S_4\Bigr)
u(p),\nonumber
\end{eqnarray}
where these functions are denoted $g_{1-4}$, and $S_{1-4}$ are
basis spin structures. It is convenient to construct the the
latter ones as follows~\cite{karm98}:
\begin{eqnarray}
\label{base1}
S_1&=&2x_1-(1+x_1) \displaystyle{\frac{m \sla
\omega}{\omega\cd p}},
\nonumber \\
S_2&=&\displaystyle{\frac{m\sla \omega}{\omega\cd p}},
\nonumber\\
S_3&=&iC_{ps}\left[2x_1 -(1-x_1)\displaystyle{\frac{m\sla
\omega}{\omega\cd p}}\right]\gamma_5,
\nonumber\\
S_4&=&i\,C_{ps}\displaystyle{\frac{m\sla \omega}{\omega\cd
p}}\gamma_5
\end{eqnarray}
with $x_1=\frac{\omega \cd k_1}{\omega \cd p}$, while
$C_{ps}$ is the following pseudoscalar:
\begin{equation}
\label{wf6}
C_{ps}=\frac{1}{m^2\omega\cd p}
e^{\mu\nu\rho\gamma}k_{2\mu}k_{3\nu}p_{\rho}\omega_{\gamma}.
\end{equation}
The function $C_{ps}$ can only be constructed with four
independent four-vectors. This is the case in LFD for $n \ge 3$.
In the nonrelativistic limit, one would need $n \ge 4$. We can
then construct two additional spin structures $S_3$ and $S_4$ of
the same parity as $S_1$ and $S_2$ by combining $C_{ps}$  with
parity negative matrices constructed from $S_1$, $S_2$, and
$\gamma_5$ matrices.

\begin{figure}[bt]
\begin{center}
\includegraphics[width=20pc]{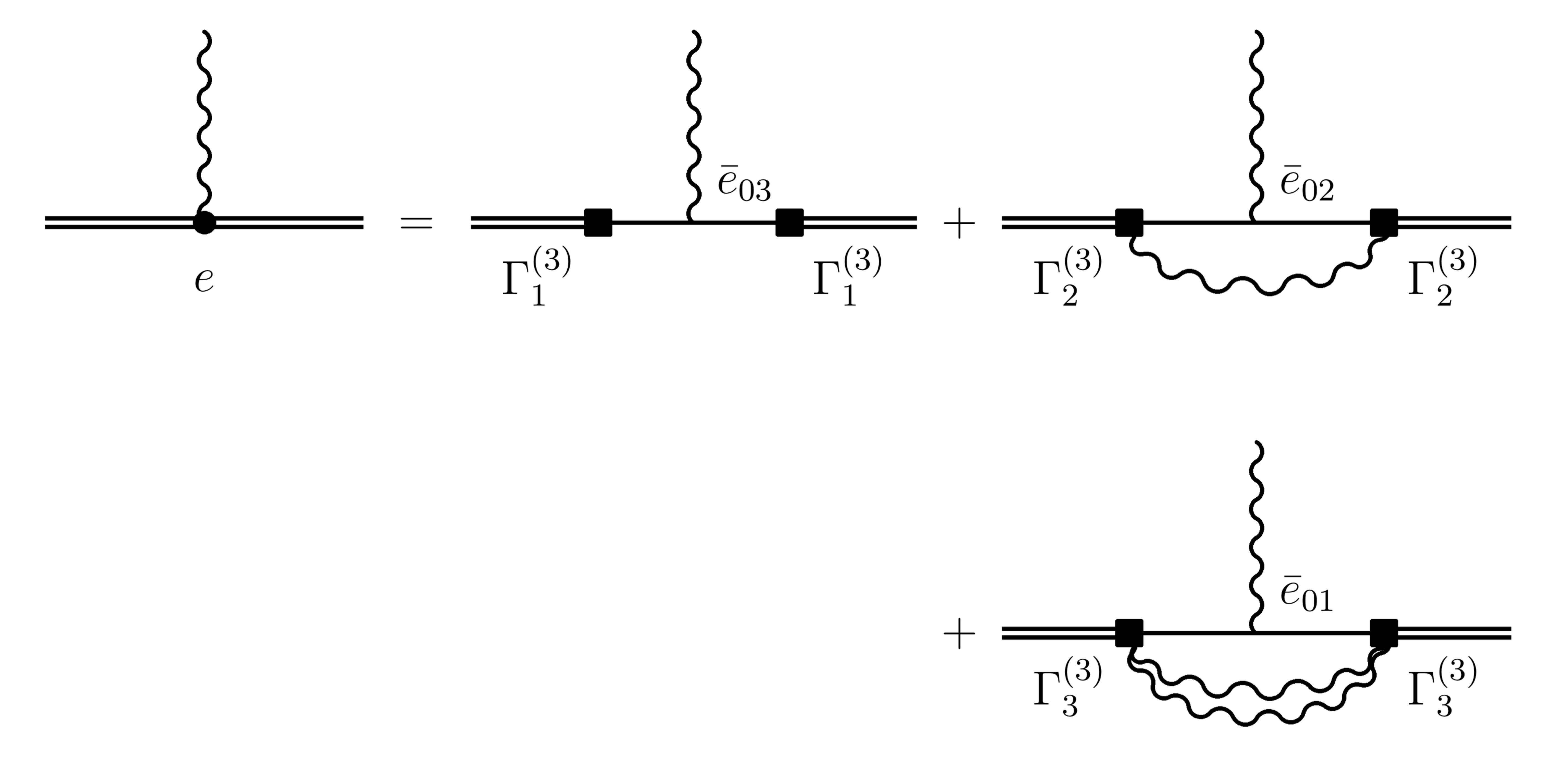}
\end{center}
\caption{Fermion-boson 3PGF in the three-body approximation.}\label{ffm3}
\end{figure}

In our computational procedure we take the limit $m_1\to\infty$
analytically and then study the limit $\mu_1\to\infty$
numerically. As a result, the calculated vertex functions depend
parametrically on $\mu_1$. The main question we are interested in
concerns the behavior of observables as a function of $\mu_1$: do
they remain finite and physically reasonable at $\mu_1\to \infty$
or not? For this purpose, we calculate the fermion anomalous
magnetic moment, using the state vector~(\ref{eq4-5}). It can be
extracted from the fermion-boson 3PGF in the three-body
approximation, as shown in Fig.~\ref{ffm3}. The detail of the
calculation can be found in~\cite{yukawa}. We just recall here the
main numerical results.

The anomalous magnetic moment is calculated for a typical set of
physical parameters $m=0.938$ GeV, $\mu = 0.138$ GeV, and two
values of the coupling constant $\alpha \equiv
\frac{g^2}{4\pi}=0.2$ and $0.5$. This mimics, to some extent, a
physical nucleon coupled to scalar "pions". The typical
pion-nucleon coupling constant is given by
$g=\frac{g_A}{2F_\pi}\langle k\rangle$ where $\langle k\rangle$ is
a typical momentum scale, and $g_A$ and $F_\pi$ are the axial
coupling constant and the pion decay constant, respectively. For
$\langle k\rangle = 0.2$ GeV we just get $\alpha \simeq 0.2$.
\begin{figure}[ht!]
\begin{center}
\includegraphics[width=8.5cm]{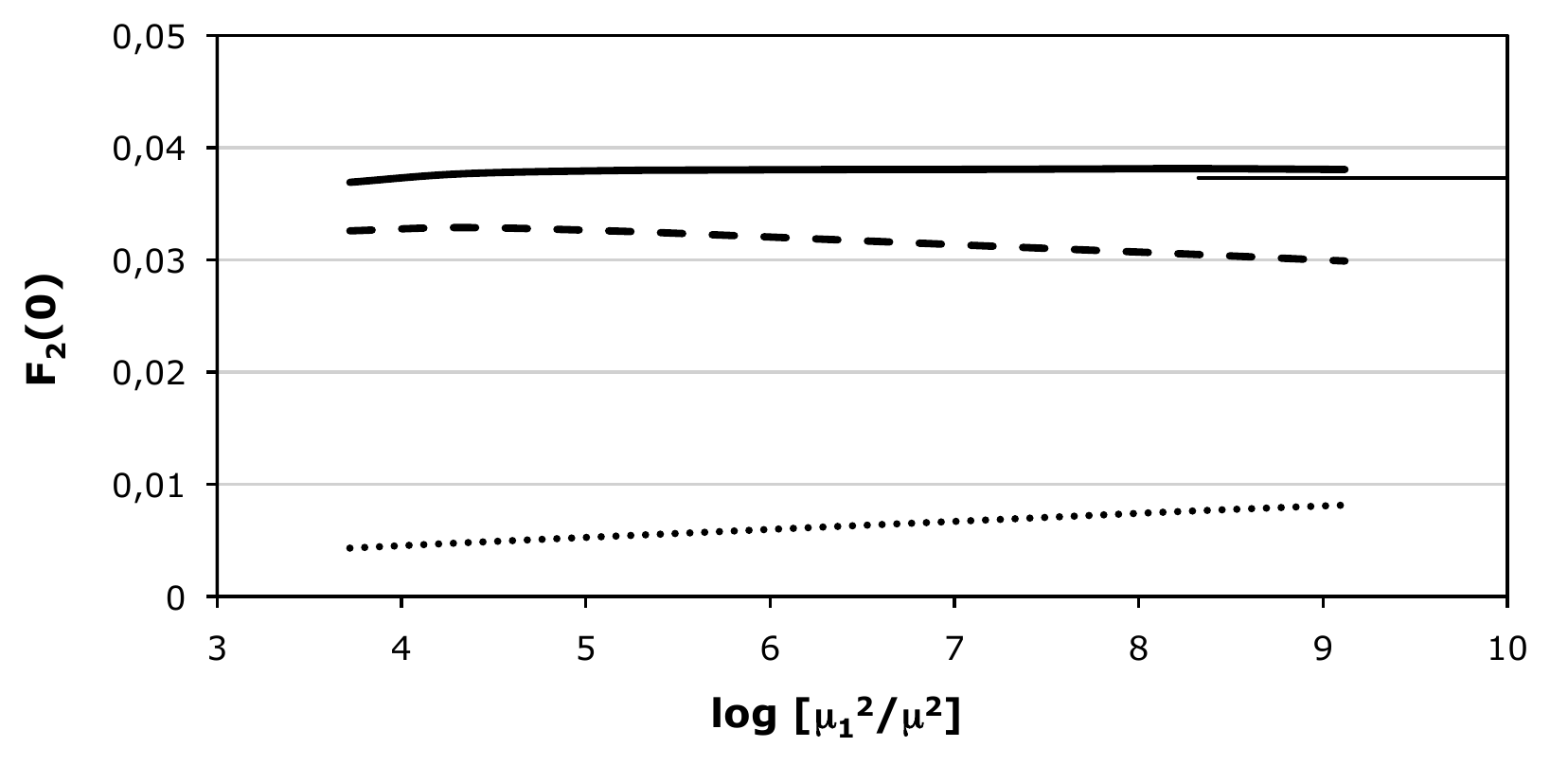}
\includegraphics[width=8.5cm]{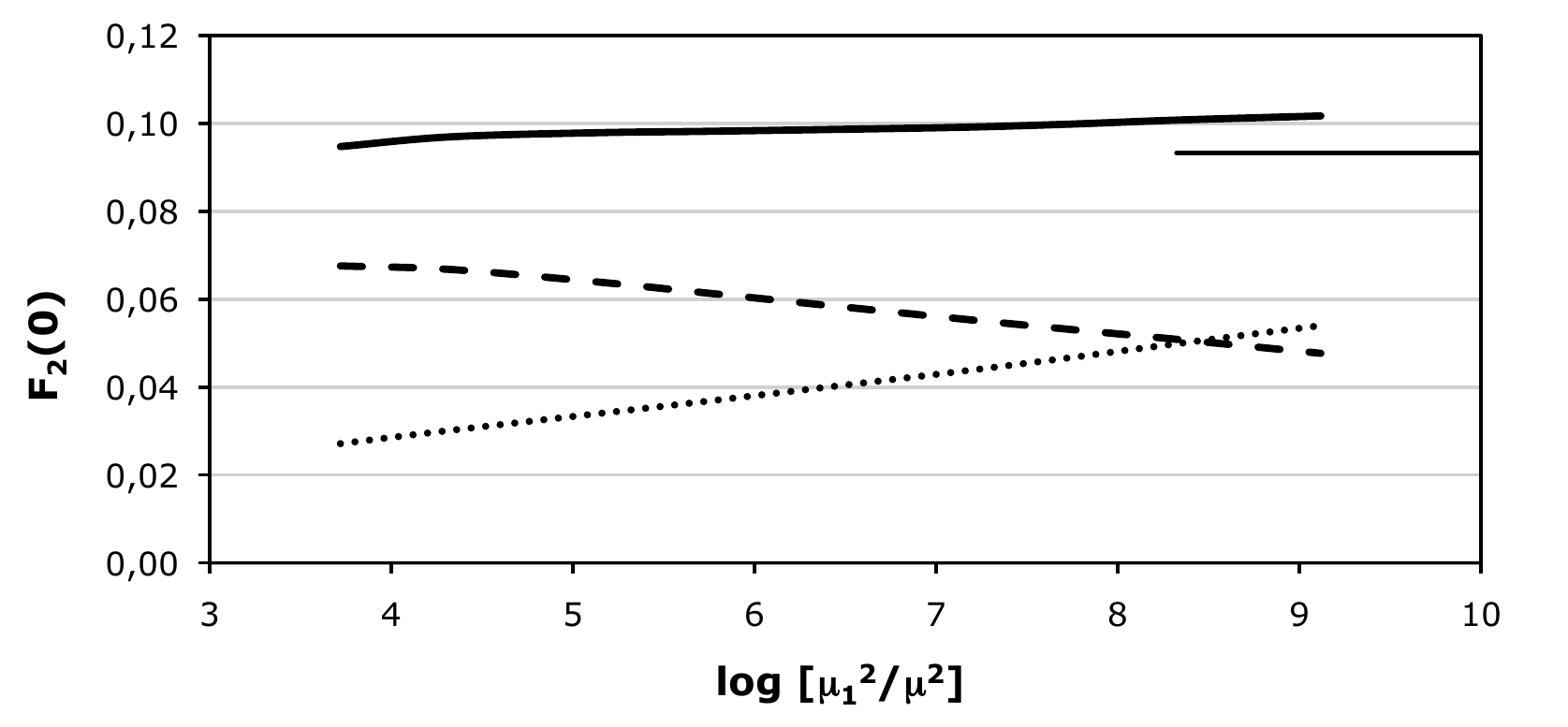}
\end{center}
\caption{The anomalous magnetic moment in the Yukawa model as a
function of the PV mass $\mu_1$, for two different values of the
coupling constant, $\alpha = 0.2$ (upper plot) and $0.5$ (lower
plot). The dashed and dotted lines are, respectively, the two- and
three-body contributions, while the solid line is the total
result. The value of the anomalous magnetic moment calculated in
the $N=2$ approximation is shown by the thin line on the right
axis.}\label{amm}
\end{figure}

We plot in Fig.~\ref{amm} the anomalous magnetic moment as a
function of $\log\left[\frac{\mu_1^2}{\mu^2}\right]$, for the two
different values of $\alpha$ mentioned above. We show also on each
of these plots the value of the anomalous magnetic moment
calculated for the $N=2$ truncation, which coincides with the
anomalous magnetic moment obtained in the second order of
perturbation theory. The results for $\alpha = 0.2$ show rather
good convergence as $\mu_1\to \infty$. The contribution of the
three-body Fock sector to the anomalous magnetic moment is
sizeable but small, indicating that the Fock
decomposition~(\ref{Fock}) converges rapidly. This may show that
once higher Fock components are small, we can achieve a
practically converging calculation of the anomalous magnetic
moment. Note that this value of $\alpha$ is not particularly
small: it is about 30 times the electromagnetic coupling, and is
about the size of the typical pion-nucleon coupling in  a nucleus.

When $\alpha$  increases, we see that the contribution of the
three-body sector considerably increases. For $\alpha = 0.5$  the
three-body contribution to the anomalous magnetic moment starts to dominate at large
values of $\mu_1$. The dependence of the anomalous magnetic moment on the PV boson mass
$\mu_1$ becomes more appreciable,
although it keeps rather small.
\begin{figure}[ht!]
\begin{center}
\includegraphics[width=8.5cm]{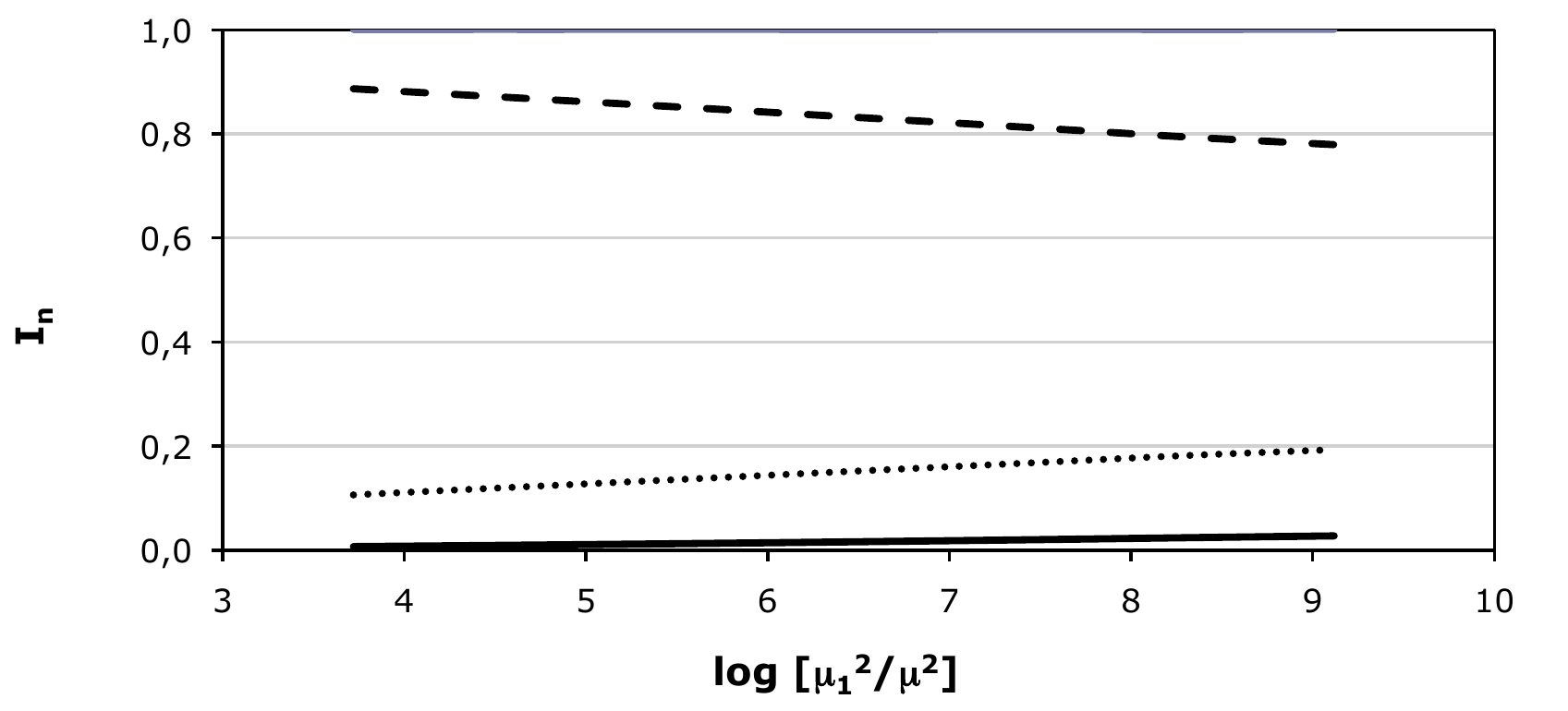}
\includegraphics[width=8.5cm]{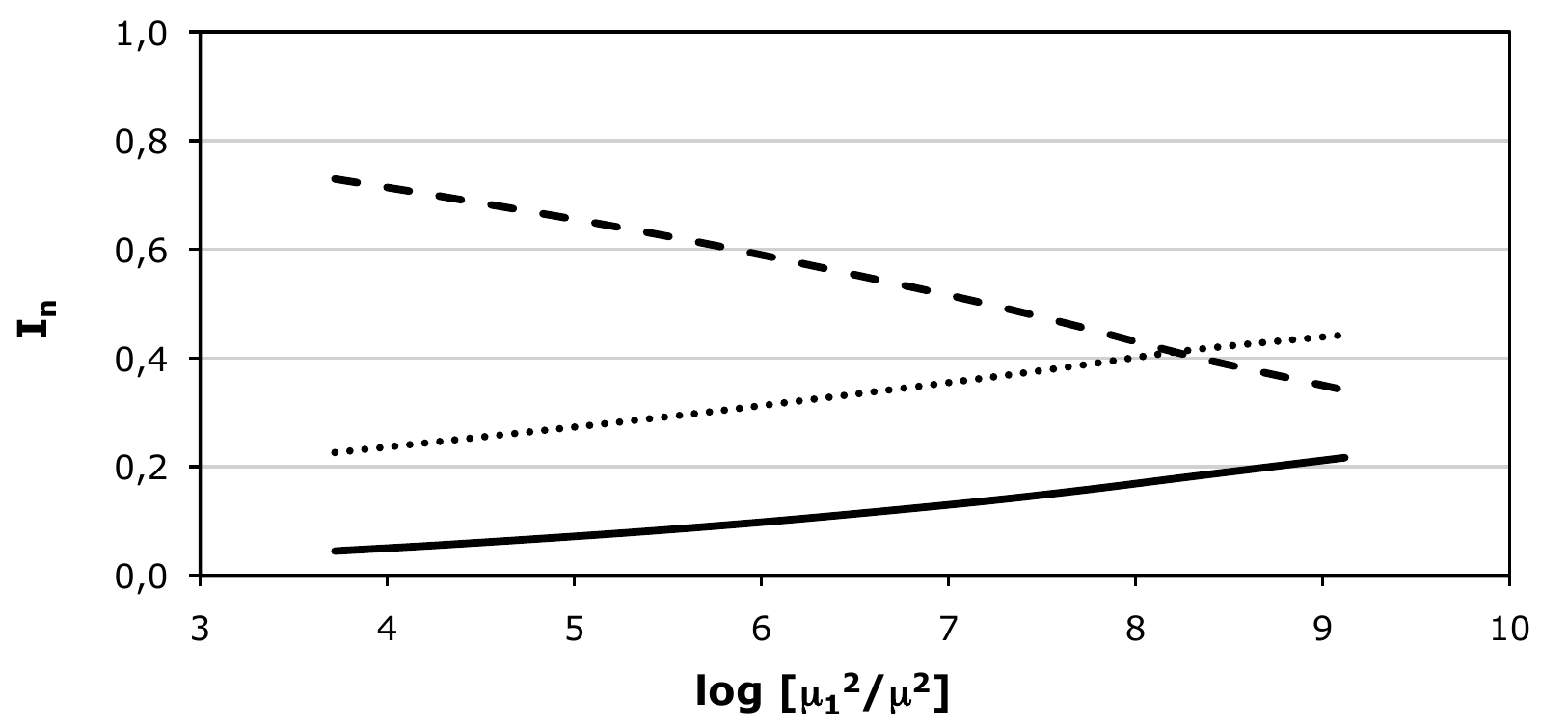}
\end{center}
\caption{Individual contributions of the one- (dashed line), two-
(dotted line), and three-body (solid line) Fock sectors to the
norm, fixed to $1$, of the state vector, as a
function of the PV boson mass $\mu_1$, for $\alpha =0.2$ (upper
plot) and $\alpha = 0.5$ (lower plot).}\label{n123}
\end{figure}

In order to have a more physical insight into the relative
importance of different Fock sectors in the
decomposition~(\ref{Fock}) for the state vector, we plot in
Fig.~\ref{n123} the contributions of the one-, two-, and
three-body Fock sectors to the norm of the state vector for the
two values of the coupling constant, considered in this work. We
see again that at $\alpha=0.2$ the three-body contribution to the
norm is small, while it is not negligible and increases with
$\mu_1$, when $\alpha = 0.5$.

According to renormalization theory, the PV boson mass should be
much larger than any intrinsic momentum scale present in the
calculation of physical observables. With this limitation,
physical observables should be independent of any variation of the
PV boson mass, within an accuracy which can be increased at will.
This is what we found in our numerical calculation for small
enough values of $\alpha$.
\begin{figure}[ht!]
\begin{center}
\includegraphics[width=8.5cm]{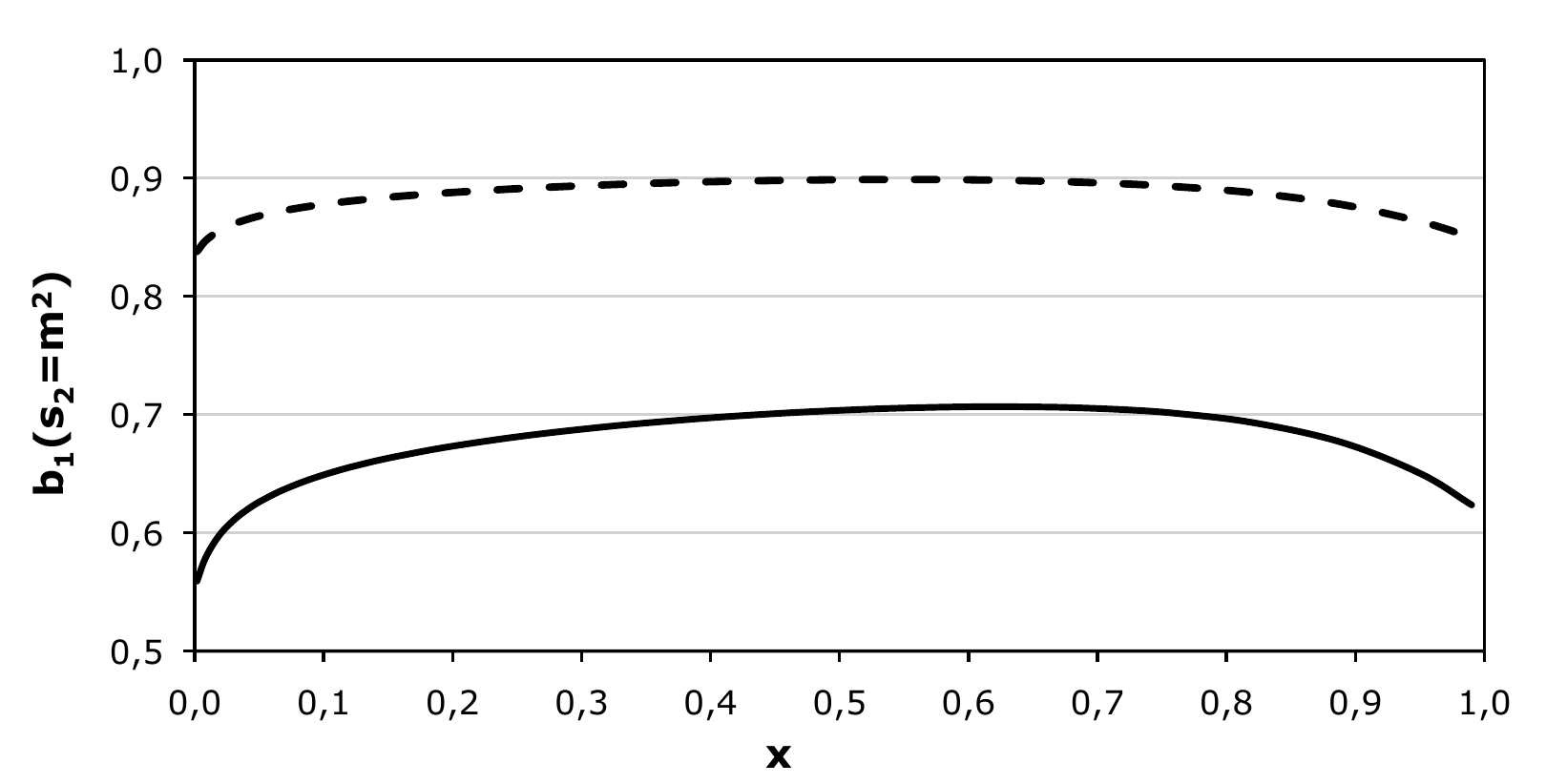}
\includegraphics[width=8.5cm]{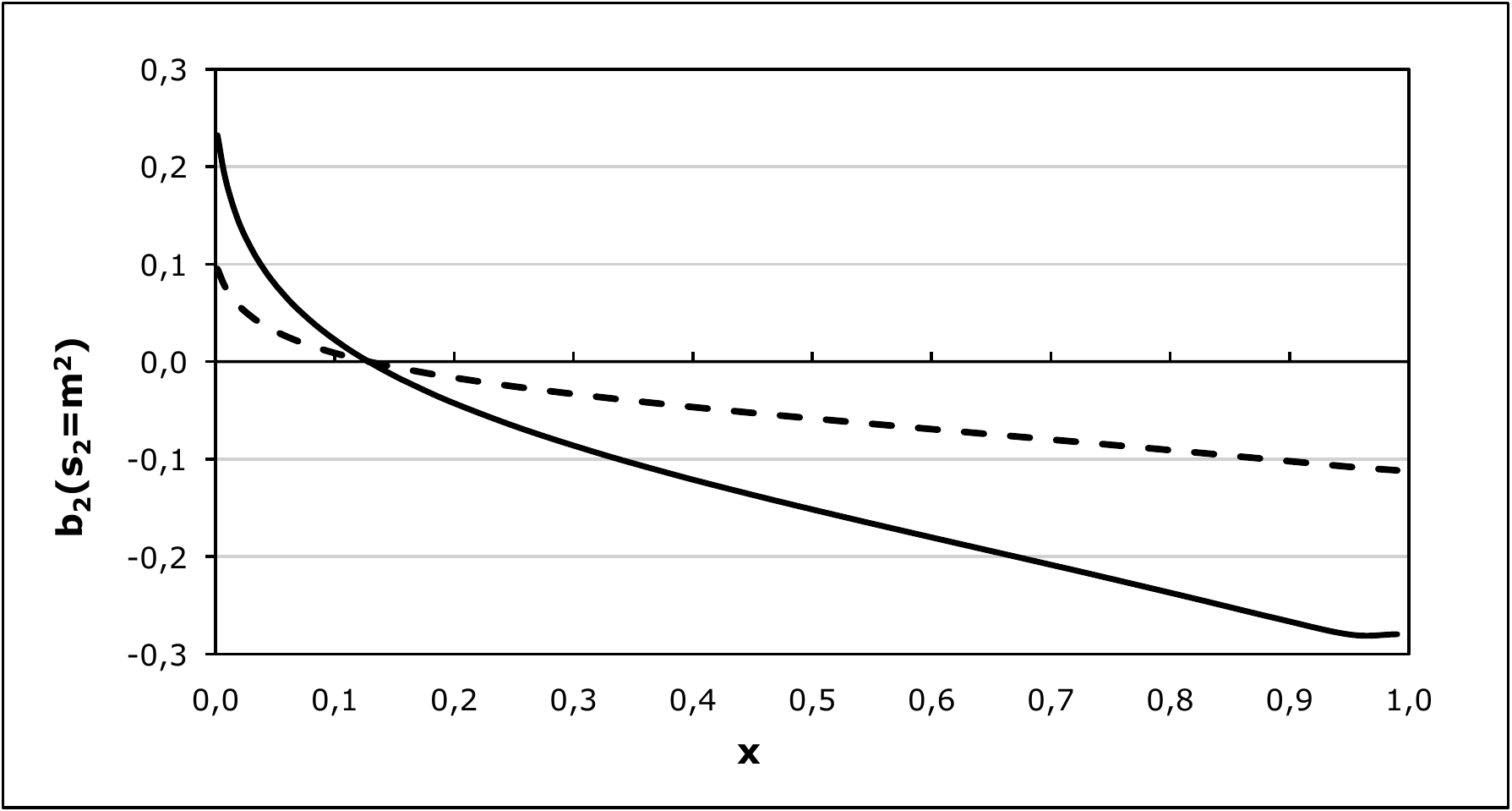}
\end{center}
\caption{The spin components $b_1$ (upper plot) and $b_2$ (lower
plot) of the two-body vertex function calculated at $s_2=m^2$, as
a function of $x$, for $\alpha = 0.2$ (dashed line) and $\alpha =
0.5$ (solid line), for a typical value of $\mu_1=100$
GeV.}\label{b1on}
\end{figure}

In order to understand the possible origin of the residual
dependence of the anomalous magnetic moment on $\mu_1$, we plot in
Figs.~\ref{b1on} the two-body spin components $b_1$ and
$b_2$ calculated at $s_2=m^2$, as a function of $x$. As we
already mentioned in Sec.~\ref{renorcond}, the spin components for
physical particles $b_1(s_2=m^2)$ and $b_2(s_2=m^2)$
should be independent of $x$ in an exact calculation. Moreover,
$b_2(s_2=m^2)$ should be zero. It is here fixed to zero at a
given value of $x=x^*\equiv\frac{\mu}{m+\mu}$, by the adjustment
of the constant $Z_\omega$. We clearly see in these plots that the
on-shell $b_1$ is not a constant, although its dependence on
$x$ is always weak, while $b_2$ is not identically zero,
although its value is relatively smaller than that of $b_1$
for $\alpha=0.2$, and starts to be not negligible for $\alpha =
0.5$. A similar situation is observed in the scalar case, when we
remove "by hands", from the state vector, the Fock sector with the
antiparticle (see the dash-dotted curve in Fig.~\ref{G2x_st}).

It is instructive to study the properties of
$b_{1,2}(s_2=m^2)$ in perturbation theory. This can be done
by calculating the amplitudes of the diagrams shown in
Fig.~\ref{anti} (but for fermions, of course!). Note that
in our Fock space truncation~(\ref{eq4-5}), the contribution of the
$Fbb$ intermediate state (the left diagram in Fig.~\ref{anti}) is
automatically taken into account by the solution of the eigenstate
equation, while the contribution of the $FF\bar{F}$ state (the
right diagram in the same figure) is absent. If one calculates the
sum of both contributions within perturbation theory~\cite{kms09},
one finds
\begin{eqnarray}
b_{1}^{pert}(s_2=m^2)&=& const, \\
b_{2}^{pert}(s_2=m^2)&=& 0.
\end{eqnarray}
This is a first indication that the expected properties of the
on-shell functions $b_{1,2}$ are indeed recovered, when
antifermion degrees of freedom are involved. This is also a
confirmation of similar features found in Sec.~\ref{scalar} for
the scalar system.

\subsection{Light-front chiral effective field theory} \label{LFCEFT}
The calculation of baryon properties within the framework of
chiral perturbation theory is a subject of active theoretical
developments. Since the nucleon mass is not zero in the chiral
limit, all momentum scales are {\em a priori} involved in the
calculation of baryon properties (like masses or electro-weak
observables) beyond tree level. This is at variance with the meson
sector for which a meaningful power expansion of any physical
amplitude can be done.

While there is not much freedom, thanks to chiral symmetry, for
the construction of the effective Lagrangian in chiral
perturbation theory in terms of the pion field --- or more
precisely in terms of the U field defined by $U=e^{i {\bf
\tau}.{\bf \pi}/f_\pi}$, where $f_\pi$ is the pion decay constant
and $\tau$ are the Pauli matrices, --- one should settle an
appropriate approximation scheme in order to calculate baryon
properties. Up to now, two main strategies have been adopted. The
first one is to force the bare (and hence the physical) nucleon
mass to be infinite, in heavy baryon chiral perturbation
theory~\cite{manohar}. In this case, by construction, an expansion
in characteristic momenta can be developed. The second one is to
use a specific regularization scheme~\cite{IR} in order to
separate contributions which exhibit a meaningful power expansion,
and hide the other parts in appropriate counterterms. In both
cases however, the explicit calculation of baryon properties
relies on an extra approximation in the sense that physical
amplitudes are further calculated by expanding the effective
Lagrangian, denoted by ${\cal L}_{CPT}$, in a finite number of
pion fields.

Moreover, it has recently been realized that the contribution of
pion-nucleon resonances, like the $\Delta$ and Roper resonances,
may play an important role in the understanding of the nucleon
properties at low energies~\cite{delta}. These resonances are just
added "by hand" in the chiral effective Lagrangian. This is also
the case for the most important $2\pi$ resonances, like the
$\sigma$ and $\rho$ resonances.

\begin{figure}[btph]
\begin{center}
\includegraphics[width=18pc]{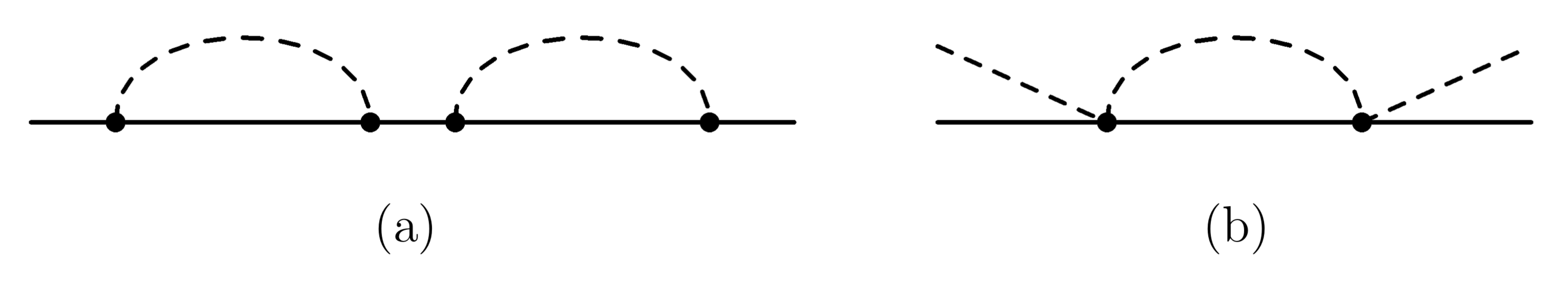}
\caption{Iteration of the self-energy contribution in the first
order of perturbation theory (a); irreducible contribution to the
bound state equation (b). Dashed lines represent
pions.\label{self_gen}}
\end{center}
\end{figure}

Since in the chiral limit the pion mass is zero, any calculation
of $\pi N$ systems demands a relativistic framework to get, for
instance, the  right analytical properties of the physical
amplitudes. The calculation of compound systems, like a physical
nucleon composed of a bare nucleon coupled to many pions, relies
also on a nonperturbative eigenstate equation. While the mass of
the system can be determined in leading order from the iteration
of the $\pi N$ self-energy calculated in the first order of
perturbation theory, as indicated in Fig.~\ref{self_gen}(a), this
is in general not possible, in particular, for $\pi N$ irreducible
contributions, as shown in Fig.~\ref{self_gen}(b).

\begin{figure}[btph]
\begin{center}
\includegraphics[width=18pc]{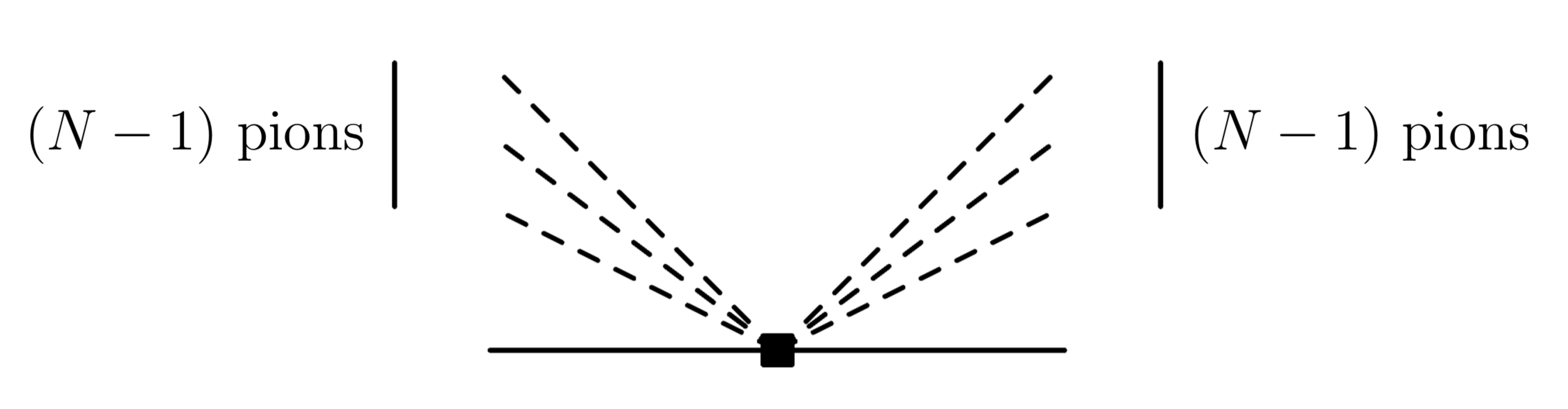}
\caption{General vertex including a maximum of $(N-1)$ pion fields
in the initial and final states. \label{counting}}
\end{center}
\end{figure}

The general framework we have developed above is particularly
suited to deal with these requirements. This leads to the
formulation of light-front chiral effective field theory
(LF$\chi$EFT)~\cite{LF_jf} with a specific effective Lagrangian
${\cal L}_{eff}$. The decomposition of the state vector in a
finite number of Fock components implies to consider an effective
Lagrangian which enables all possible elementary couplings between
the pion and nucleon fields compatible with the Fock space
truncation. This is indeed easy to achieve in chiral perturbation
theory, since each derivative of the U field involves one
derivative of the pion field. In the chiral limit, the chiral
effective Lagrangian of order $p$, ${\cal L}_{CPT}^p$, involves
$p$ derivatives and therefore at least $p$ degrees of the pion
field. In order to calculate the state vector in the $N$-body
approximation, with one fermion and $(N-1)$ pions, one has
therefore to include contributions up to $2(N-1)$ pion fields in
the effective Lagrangian, as shown in Fig.~\ref{counting}. We thus
should calculate the state vector in the $N$-body approximation
with an effective Lagrangian denoted by ${\cal L}_{eff}^N$ and
given by
\begin{equation}
{\cal L}_{eff}^N={\cal L}_{CPT}^{p=2(N-1)}\ .
\end{equation}

While the effective Lagrangian in LF$\chi$EFT can be mapped out to
the CPT Lagrangian of order $p$, the calculation of the state
vector does not rely on any momentum decomposition. It relies only
on an expansion in the number of pions in flight at a given
light-front time. In other words, it relies on an expansion  in
the fluctuation time, $\tau_f$, of such a contribution. From
general arguments, the more particles we have at a given
light-front time, the smaller the fluctuation time is. At low
energies, when all processes have characteristic interaction times
larger than $\tau_f$, this expansion should be meaningful.

\begin{figure}[btph]
\begin{center}
\includegraphics[width=15pc]{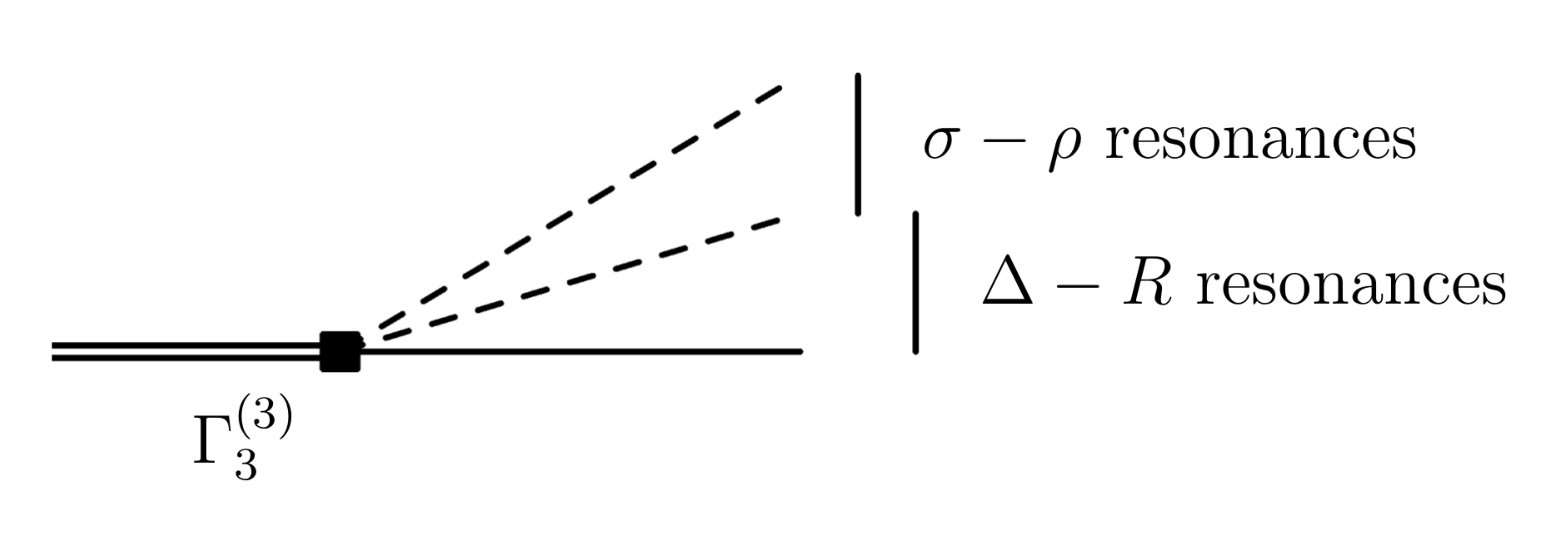}
\caption{Three-body vertex function which exhibits both the contribution of $2\pi$
resonances and $\Delta$ as well as Roper nucleon resonances. \label{delta}}
\end{center}
\end{figure}

It is interesting to illustrate the general features of
LF$\chi$EFT calculations. At order $N=2$, we already have to deal
with irreducible contributions, as shown in
Fig.~\ref{self_gen}(b). It leads to non-trivial renormalization of
the coupling constant. The calculation at order $N=3$ incorporates
explicitly contributions coming from $\pi \pi$ interactions, as
well as all low energy $\pi N$ resonances, like the $\Delta$ or
Roper resonances, as shown schematically in Fig.~(\ref{delta}).
Indeed, in the $\vert \pi \pi N\rangle$ Fock sector, the $\pi N$
state can couple to both $J=T=3/2$ as well as $J=T=1/2$ states. We
can generate therefore all $\pi N$ resonances in the intermediate
state without the need to include them explicitly, provided the
effective Lagrangian has the right dynamics to generate these
resonances. This is the case, by the construction, in CPT.

Preliminary results obtained by using the PV regularization scheme
can be found in~\cite{LF_kmt}.

%
\section{Conclusion and perspectives}
The understanding of the properties of relativistic compound systems from
an elementary Hamiltonian in nuclear and particle physics demands
to develop a nonperturbative framework. This framework should
include a well-defined strategy for approximate calculations of
these properties and a systematic way to improve the accuracy.

We have described
in this review a general framework based on light-front dynamics.
In this scheme, the state vector of any system of interacting
particles is decomposed in Fock components. Since for obvious
practical reasons this decomposition
should be
truncated to take into account a finite number of Fock components,
we have shown how to control in a systematic way the convergence
of such expansion.

Our formalism relies, first, on the covariant formulation of
light-front dynamics, and, second, on a systematic nonperturbative
renormalization scheme in order to avoid any uncancelled
divergences. The applications we have presented on QED, on a purely scalar
system, on the Yukawa model, and on chiral effective field
theory on the light-front have shown the flexibility, the real
advantages, and the nice features of our formalism.

Several developments will soon achieve to settle a complete
framework to deal with field theory on the light front. This
includes first the full account of antiparticle degrees of freedom
in order to recover, order by order in the Fock expansion, the
scale invariance of physical observables for arbitrary values of
the coupling constant. This scale invariance should be checked
within a given regularization scheme.

We have used up to now the Pauli-Villars regularization scheme.
While this scheme is systematic and can be applied to a variety of
physical systems, it may be cumbersome to implement from a
numerical point of view in higher order calculations, since it
involves many non-physical components including Pauli-Villars
fields. It also demands to perform calculations with very large
mass scales (the Pauli-Villars fermion and boson masses). The use
of the Taylor-Lagrange renormalization scheme~\cite{TLRS} has
proven  to be a very natural scheme in light-front dynamics. It
should be applied to more involved calculations.

Finally, it is now well-recognized that the understanding of
spontaneous symmetry breaking phenomena on the light-front can be
achieved by taking into account nonperturbative zero mode
contributions to field operators~\cite{hksw}. It remains to
include these zero modes in the general framework we developed in
this review.

\begin{acknowledgements}
Three of us (V.A.K., A.V.S. and N.A.T.) are sincerely grateful for
the warm hospitality of the Laboratoire de Physique Corpusculaire,
Universit\'e Blaise Pascal, in Clermont-Ferrand, where a part of
the present study was performed. This work has been supported by
grants from CNRS/IN2P3 and the Russian Academy of Science.
\end{acknowledgements}


\end{document}